\documentclass[12pt]{article}
\usepackage{eurosym}
\usepackage{float}
\usepackage{graphicx}
\usepackage[utf8]{inputenc}
\usepackage[T1]{fontenc}
\usepackage{indentfirst}
\usepackage[margin=0.6in,nomarginpar]{geometry}
\usepackage[final]{hyperref}
\usepackage{amsmath}
\usepackage{hyperref}
\usepackage{cite}
\usepackage{subcaption}
\usepackage{caption}
\usepackage{amssymb}
\usepackage{multirow}
\usepackage[table]{xcolor}
\usepackage{orcidlink}
\usepackage{color}
\usepackage{hyperref}

\usepackage[utf8]{inputenc}

\hypersetup{
colorlinks=true,
linkcolor=blue,
citecolor=blue,
filecolor=magenta,
urlcolor=blue
}

\begin{document}

\title{Phantom RN-AdS black holes in noncommutative space}
\author{B. Hamil \orcidlink{0000-0002-7043-6104} \thanks{%
hamilbilel@gmail.com}  \\
Laboratoire de Physique Math\'{e}matique et Subatomique,\\
Facult\'{e} des Sciences Exactes, Universit\'{e} Constantine 1, Constantine,
Algeria. \and B. C. L\"{u}tf\"{u}o\u{g}lu 
\orcidlink{Orcid ID :
0000-0001-6467-5005} \thanks{%
bekir.lutfuoglu@uhk.cz (Corresponding author)} \\
Department of Physics, Faculty of Science, University of Hradec Kralove, \\
Rokitanskeho 62/26, Hradec Kralove, 500 03, Czech Republic. }
\date{\today }
\maketitle

\begin{abstract}
We analyze the effects of noncommutativity on phantom Reissner-Nordström-Anti-de Sitter black holes by modeling mass and charge distributions with Lorentzian profiles. The modified metric function exhibits significant deviations from the classical case, leading to changes in the horizon structure and the suppression of singularities. Through a comparative thermodynamic analysis, we derive expressions for the mass, Hawking temperature, entropy, and heat capacity, identifying stability conditions and phase transitions induced by noncommutative corrections. The efficiency of the black hole as a heat engine is evaluated, showing that noncommutativity influences the thermodynamic cycle differently in the presence of phantom fields. Furthermore, we investigate the orbital motion of test particles and photons, deriving the effective potential, innermost stable circular orbits, and the shadow profile. Finally, we compute quasinormal modes to assess dynamical stability, revealing that noncommutativity modifies the damping behavior and introduces a new branch of non-oscillatory modes, absent in the classical case. Our findings provide a deeper understanding of the interplay between phantom fields, noncommutative geometry, and black hole thermodynamics, offering potential observational signatures for exotic compact objects.  
\end{abstract}

\section{Introduction}

Black holes are among the most enigmatic objects in theoretical physics, arising from the solutions to Einstein's equations in General Relativity. These compact regions of spacetime exhibit an event horizon beyond which nothing, not even light, can escape. Black holes play a pivotal role in our understanding of gravitational phenomena and provide a unique laboratory for exploring fundamental concepts of spacetime, singularity formation, and quantum gravity. Among the various classes of black holes, Reissner-Nordström (RN) black holes, which include both electric charge and gravitational effects, are especially important in the study of electromagnetic interactions and their influence on spacetime geometry \cite{Reissner, Weyl, Nordstrom, Jeffery}. {When coupled with a negative cosmological constant, these black holes take the form of Reissner-Nordström-Anti-de Sitter (RN-AdS) black holes. In AdS space, the structure is different from flat space, featuring a cosmological horizon and a well-defined timelike boundary. There is no event horizon in the conventional sense as AdS space is hyperbolic and extends infinitely} \cite{Louko1996}. RN-AdS black holes have garnered significant attention due to their relevance in high-energy physics, particularly in the context of holography, string theory, and AdS/CFT correspondence \cite{Witten1998, Chamblin1999, Maldacena1999}, where they provide insight into the structure of spacetime at the boundary and offer possible models for understanding quantum gravity in curved spacetime \cite{Wang2000}.

Black holes are thermodynamic objects, possessing properties such as temperature, entropy, and heat capacity \cite{Bekenstein1973, Hawking1975}, governed by the laws of black hole thermodynamics \cite{Bardeen1973}. These laws, analogous to the classical laws of thermodynamics, suggest deep connections between gravity, quantum mechanics, and statistical physics. In the context of AdS black hole thermodynamics, a richer structure emerges \cite{Hawking1983, Brown1994, Cai1996, Peca1999}. In particular, interpreting the cosmological constant as thermodynamic pressure allows for the exploration of thermodynamic volume and phase transitions \cite{Dolan2011, Kubiznak2012, Kubiznak2017, Hamil20241}. For RN-AdS black holes, this perspective offers deeper insight into their stability, phase transitions, critical behavior, and the analogy with the Van der Waals fluids \cite{Kooplya2008, Niu2012, Liu2014, Chen2019, Li2020, Zhang2021, Ghaffarnejad2025}.

Despite their extreme nature, black holes cannot be directly observed, as they do not emit light; however, their presence can be inferred through various observational signatures. One such signature is the black hole shadow, a dark region cast by the event horizon against the backdrop of luminous matter. The concept of black hole shadows has gained prominence due to recent breakthroughs, particularly with the Event Horizon Telescope (EHT), which provided the first-ever image of a black hole shadow for M87* in 2019 and later for Sagittarius A*, the supermassive black hole at the center of the Milky Way \cite{M871, SagA1}. These observations confirm general relativity’s predictions and offer a means to probe deviations from Einstein’s theory and explore alternative gravitational models. The shadow’s size and shape help infer properties such as mass, spin, and the surrounding spacetime geometry.  Accompanying the astrophysical observations, theoretical studies have investigated shadows in various spacetimes, including charged and AdS black holes \cite{Atamurotov2013, Roman2019, Guo2020, Anacleto2021, Pantig2022, Okyay2022, Hamil20231, Hamil20232, Hamil20243, Yunusov2024, Luo2024, Hamil20251, Hamil20252, Hamil20253}. In particular, studies of RN-AdS black hole shadows have explored how charge and AdS curvature affect their structure, uncovering connections between black hole thermodynamics, phase transitions, and optical properties \cite{Belhaj2020, Wang2022, Mandal2023, Ladino2024, AraujoFilho2025}. These investigations suggest that variations in these parameters can leave observational imprints on the shadow’s size and shape, offering a potential avenue for testing alternative gravity models.

Beyond black hole shadows, another fundamental signature of black holes lies in their response to external perturbations, encoded in quasinormal modes (QNMs). { The implication that all oscillations decay over time inherently suggests stability, whereas the key point is that in some cases, the oscillations grow, indicating instability}\cite{Kokkotas:1999bd, Berti:2009kk, Konoplya:2011qq}. Mathematically, QNMs are represented by complex frequencies, where the real part determines the oscillation frequency and the imaginary part governs the damping rate \cite{Konoplya2002a}. Although QNMs have been studied for decades \cite{Press1971, Aharony:1999ti, Starinets:2002br, Kovtun:2005ev, Konoplya:2007jv, Dyatlov:2010hq, Dyatlov:2011jd, Konoplya:2017zwo}, recent gravitational wave observations—such as those from LIGO, Virgo, and the upcoming LISA mission \cite{BF1, BF2, BF3}—have renewed interest in the field, driving further theoretical advancements \cite{Dubinsky:2024gwo, Bolokhov:2023bwm, Konoplya:2022pbc, Luo:2024dxl, Zinhailo:2024kbq}. Extensive studies have explored QNMs in AdS black holes across various contexts \cite{Horowitz:1999jd, Konoplya:2002zu, Wang:2004bv, Cardoso:2003cj, Konoplya:2017ymp, Konoplya:2022hll, Fontana2022, Konoplya:2023kem, Ficek2024, Lin2024, Guo2024}. Within particular focus on RN-AdS black holes,  Wang et al. investigated massless scalar field perturbations, confirming that in the near-extremal limit, the imaginary part of the QNM frequency vanishes, leading to power-law decay at late times \cite{Wang:2004bv}. Later, Fontana et al. revealed a fine structure in the quasinormal spectrum of accelerating AdS black holes, distinguishing them from their non-accelerating counterparts and confirming their stability under linear scalar perturbations \cite{Fontana2022}. More recently, Ficek et al. examined the quasinormal spectrum near extremality, uncovering a transition where the least damped mode ceases to oscillate, fundamentally altering the black hole’s late-time response \cite{Ficek2024}.

In recent years, interest has grown in understanding the influence of exotic matter fields on black hole configurations. Phantom fields, defined by their negative energy density and violation of the weak energy condition, have been a topic of considerable intrigue \cite{Caldwell2002}. These fields arise in various theoretical contexts, including models of dark energy and modifications to general relativity \cite{Nojiri2003, Elizalde2004}. When coupled to black holes, phantom fields lead to intriguing phenomena, such as modifications to event horizons, altered causal structures, non-trivial spacetime topologies \cite{Bronnikov2006, Clement2009, Bronnikov2012}. In particular, coupling phantom fields to RN black holes significantly modifies their thermal quantities \cite{Jamil2011} and QNMs \cite{Pan2011}. When extended to RN-AdS black holes, these modifications become even more pronounced, introducing substantial alterations to both geometric and thermodynamic properties. The negative energy contribution reshapes the horizon structure, often resulting in a single event horizon instead of the typical two seen in classical charged black holes, further influencing their causal and thermodynamic behavior \cite{Jardim2012}. Thermodynamically, these configurations exhibit unique behaviors, such as the absence of $T-S$ or $P-V$ criticality and reversed energy contributions in the first law of black hole thermodynamics, leading to decreased mass-energy of the system \cite{Quevedo2016, Mo2018, Han2020, Salah, Shahzad2024}. In extended phase space, phantom RN-AdS black holes display distinct Gibbs free energy profiles and lack first-order phase transitions, further diverging from classical RN-AdS solutions \cite{Han2020}. These changes also impact geodesic structures and causal diagrams, providing novel avenues for studying matter interactions in highly curved spacetimes. The theoretical implications extend to testing alternative thermodynamic frameworks, such as geometrothermodynamics, which reveal nuanced stability conditions despite occasional divergence from classical predictions \cite{Quevedo2016, Salah}. 


Parallel to these developments, noncommutative geometry has emerged as a promising extension to classical spacetime descriptions, rooted in quantum gravity theories \cite{Niki2009}. By introducing a fundamental length scale, noncommutative spacetime regularizes singularities and alters the structure of black hole solutions. The incorporation of noncommutative effects modifies mass and charge distributions, often modeled using Gaussian \cite{Nicolini, Tejeiro, Rahaman, Nozari} or Lorentzian \cite{Rizzo, Mehdipour, Liang, Hamil} profiles. These modifications lead to significant changes in black hole properties, such as the elimination of divergences in the Hawking temperature at small scales and the introduction of a minimum mass corresponding to the fundamental length scale, effectively preventing complete evaporation \cite{Nicolini2005, Nasseri2005}. The impact of noncommutativity is particularly pronounced for mini black holes, where spacetime fuzziness induced by the uncertainty principle becomes dominant \cite{Campos2022}. Studies have explored various thermodynamic quantities, including entropy, heat capacity, and phase transitions, revealing that noncommutative corrections contribute to entropy and alter the heat {capacity} thus the stability profile of black holes \cite{Campos2022}. These effects have been extended to include the influence of quintessence matter fields, further modifying black hole metrics and their observable properties, such as shadows and quasinormal modes, which serve as signatures of noncommutative corrections \cite{Hamil}. In the literature, Reissner-Nordström (RN) black hole solutions have been analyzed within the framework of noncommutative geometry corrections \cite{Mukherjee2008, Nozari2008, Alavi2009, Nozari2010, Ciric2025}, with particular focus on the effects of spacetime fuzziness on black hole stability \cite{Kim2008}, entropy \cite{Gangopadhyay2012}, and QNMs \cite{Ciric2018, Ciric2020}. Studies on noncommutative RN-AdS black hole solutions remain limited, with only a few works, such as Liang et al., exploring their critical behavior \cite{Liang2017}. Similarly, research on charged configurations coupled with exotic fields like phantom fields is still scarce, leaving open questions about their impact on spacetime geometry and thermodynamic stability.

In light of these findings, this work investigates phantom RN-AdS black holes within a noncommutative spacetime framework, examining how the interplay between phantom fields and noncommutativity influences spacetime geometry, thermodynamics, and observational signatures. Using Lorentzian mass and charge distributions, we derive static, spherically symmetric solutions and analyze their behavior under varying noncommutative parameters. The study explores modifications to the horizon structure, singularity regularization, and thermodynamic stability, including phase transitions and heat engine efficiency. Additionally, we examine the dynamics of test particles and photons, deriving the effective potential, stable orbits, and innermost stable circular orbits (ISCOs). The shadow profile is analyzed to assess the impact of noncommutative corrections on its size and shape, offering potential observational signatures. Finally, we compute quasinormal modes to evaluate the dynamical stability of these black holes, highlighting how noncommutativity alters the damping behavior and late-time response. These findings provide new insights into the combined effects of phantom fields and quantum corrections, contributing to the broader understanding of black hole physics and its observational implications.

The rest of this paper is structured as follows. In Section \ref{sec2}, we derive the metric describing phantom RN-AdS black holes in noncommutative spacetime, incorporating Lorentzian mass and charge distributions. The thermodynamic properties of these black holes, including the Hawking temperature, entropy, and heat capacity, are analyzed in Section \ref{sec3}, where we also explore their stability and phase transitions under noncommutative corrections. Section \ref{sec4} examines the role of phantom RN-AdS black holes as heat engines, evaluating their efficiency and comparing it with the Carnot cycle. In Section \ref{sec5}, we investigate the motion of test particles and photons in the black hole’s vicinity, analyzing geodesics, the ISCOs, and the shadow cast by the black hole. Section \ref{sec6} is devoted to the study of QNMs, where we compute their spectrum and assess the impact of noncommutativity on black hole stability. Finally, we summarize our findings and discuss potential implications in Section \ref{sec7}.

\section{Phantom RN-AdS black holes in Noncommutative Space} \label{sec2}

In this section, we derive the phantom RN black hole solutions in noncommutative spacetime in the presence of a negative cosmological constant, characterized by $\Lambda = -3/\ell^2$, where $\ell$ denotes the AdS radius. The action describing the underlying geometry is given by \cite{Hang}:
\begin{equation}
S = \int d^{4}x \sqrt{-g} \left( R + 2\Lambda + 2\eta F_{\mu \nu} F^{\mu \nu} \right). \label{eq:action}
\end{equation}
Here, the first term represents the Einstein-Hilbert action, the second term denotes the contribution of the cosmological constant and the third term accounts for the coupling with the Maxwell field ($\eta = 1$) or a phantom field of spin-1 ($\eta = -1$). 

The Einstein field equations derived from the action in Eq.~\eqref{eq:action} are:
\begin{equation}
\begin{aligned}
R_{\mu \nu} - \frac{1}{2} g_{\mu \nu} R + \Lambda g_{\mu \nu} &= 8\pi \left( T_{\mu \nu} \big|_{\text{matt}} + T_{\mu \nu} \big|_{\text{el}} \right), \\
\frac{1}{\sqrt{-g}} \partial_{\mu} \left( \sqrt{-g} F^{\mu \nu} \right) &= J^{\nu},
\end{aligned}
\end{equation}
where $T_{\mu \nu} \big|_{\text{matt}}$ represents the stress-energy tensor for the matter fields, and $T_{\mu \nu} \big|_{\text{el}}$ denotes the electromagnetic stress-energy tensor.

The matter stress-energy tensor, following the formalism in \cite{Nicolini}, is expressed as:
\begin{equation}
T_{\mu \nu} \big|_{\text{matt}} = \text{diag} \left( -\rho_{\text{matt}}(r, \Theta), \rho_{\text{matt}}(r, \Theta), p_{\theta}(r, \Theta), p_{\varphi}(r, \Theta) \right),
\end{equation}
where the pressure components satisfy:
\begin{equation}
p_{\theta} = p_{\varphi} = -\rho_{\text{matt}}(r, \Theta) + \frac{r}{2} \partial_r \rho_{\text{matt}}(r, \Theta).
\end{equation}
The electromagnetic stress-energy tensor is given by the standard expression:
\begin{equation}
T_{\mu \nu} \big|_{\text{el}} = \frac{\eta}{4\pi} \left( g^{\rho \sigma} F_{\mu \rho} F_{\nu \sigma} - \frac{1}{4} g_{\mu \nu} F^{2} \right),
\end{equation}
with $F^2 = F_{\mu \nu} F^{\mu \nu}$. The electromagnetic field tensor and the corresponding current density are defined as:
\begin{equation}
F^{\mu \nu} = E(r, \Theta) \begin{pmatrix}
0 & -1 & 0 & 0 \\
1 & 0 & 0 & 0 \\
0 & 0 & 0 & 0 \\
0 & 0 & 0 & 0
\end{pmatrix}, \quad J^{\nu} = \rho_{\text{el}}(r, \Theta) \delta_{0}^{\nu}.
\end{equation}

Now, let us explore the black hole solutions incorporating non-localized distribution functions, with a particular focus on Lorentzian distributions. While Gaussian distribution functions are widely employed in the study of noncommutative black holes \cite{Nicolini, Tejeiro, Rahaman, Nozari}, Lorentzian distributions have also been shown to yield significant insights \cite{Rizzo, Mehdipour, Liang, Hamil}. However, most investigations involving Lorentzian distributions have focused on uncharged black holes. Here, we extend these studies by deriving static, spherically symmetric, asymptotically flat solutions with Lorentzian mass and charge distributions, characterized by the noncommutative parameter $\sqrt{\Theta}$. 
\begin{equation}
\rho_{\text{matt}}(r, \Theta) = \frac{M\sqrt{\Theta}}{\pi^{3/2}(r^2 + \pi \Theta)^2},
\end{equation}
\begin{equation}
\rho_{\text{el}}(r, \Theta) = \frac{e\sqrt{\Theta}}{\pi^{3/2}(r^2 + \pi \Theta)^2}.
\end{equation}
The corresponding electric field takes the form:
\begin{equation}
E(r) = \frac{4Q}{\pi^{1/2}r^2} \left( \frac{1}{2\sqrt{\pi}} \tan^{-1}\left(\frac{r}{\sqrt{\pi}\sqrt{\Theta}}\right) - \frac{r\sqrt{\Theta}}{2(\pi \Theta + r^2)} \right),
\end{equation}
where $e = 4\pi Q$. Using these energy density expressions, we seek a static, spherically symmetric, asymptotically phantom RN-AdS black hole solution of the Einstein equations. The metric is assumed to take the form:
\begin{equation}
ds^2 = -f(r) dt^2 + \frac{1}{f(r)} dr^2 + r^2 \left(d\theta^2 + \sin^2\theta \, d\varphi^2 \right),
\end{equation}
with the metric function expressed as:
\begin{equation}
f(r) = 1 - \frac{\Lambda}{3}r^2 - F(r, \Theta).
\end{equation}
By considering the time component of Eq.~\eqref{eq:action}, we obtain the differential equation:
\begin{equation}
F'(r, \Theta) + \frac{F(r, \Theta)}{r} = \frac{8M\sqrt{\Theta}r}{\pi^{1/2}(r^2 + \pi \Theta)^2} + 16\eta \frac{Q^2}{\pi r^3} \left( \frac{1}{2\sqrt{\pi}} \tan^{-1}\left(\frac{r}{\sqrt{\pi}\sqrt{\Theta}}\right) - \frac{r\sqrt{\Theta}}{2(\pi \Theta + r^2)} \right)^2.
\end{equation}
Integrating this equation, we find:
\begin{align}
F(r, \Theta) &= \frac{c}{r} + \frac{2(\eta Q^2 - 2\pi M\sqrt{\pi \Theta})}{\pi^2 (\pi \Theta + r^2)} + \frac{2}{\pi^2 r} \left( 2\pi M + \frac{\eta Q^2}{\sqrt{\pi \Theta}} \right) \arctan\left(\frac{r}{\sqrt{\pi \Theta}}\right)  \nonumber \\ & \quad - \frac{4\eta Q^2}{\pi^2 r^2} \arctan^2\left(\frac{r}{\sqrt{\pi \Theta}}\right),
\end{align}
where $c$ is an integration constant. Consequently, the metric function becomes:
\begin{align}
f(r) &= 1 - \frac{\Lambda}{3}r^2 - \frac{c}{r} + \frac{2(\eta Q^2 - 2\pi M\sqrt{\pi \Theta})}{\pi^2 (\pi \Theta + r^2)} - \frac{2}{\pi^2 r} \left( 2\pi M + \frac{\eta Q^2}{\sqrt{\pi \Theta}} \right) \arctan\left(\frac{r}{\sqrt{\pi \Theta}}\right) \notag \\
&\quad + \frac{4\eta Q^2}{\pi^2 r^2} \arctan^2\left(\frac{r}{\sqrt{\pi \Theta}}\right).
\end{align}
To simplify, we expand the metric function to first-order noncommutative corrections, yielding:
\begin{equation}
f(r) = 1 - \frac{2M}{r} + \eta \frac{Q^2}{r^2} - \frac{\Lambda}{3}r^2 + \frac{8\sqrt{\Theta}M}{\sqrt{\pi}r^2} - \frac{4\eta\sqrt{\Theta}Q^2}{\sqrt{\pi}r^3} - \frac{\frac{\eta Q^2}{\pi^{3/2}\sqrt{\Theta}} + c}{r}. \label{eq:metric_function_intermediate}
\end{equation}
By fixing the integration constant as:
\begin{equation}
c = -\frac{\eta Q^2}{\pi^{3/2}\sqrt{\Theta}},
\end{equation}
the metric function simplifies to:
\begin{equation}
f(r) = 1 - \frac{2M}{r} + \eta \frac{Q^2}{r^2} - \frac{\Lambda}{3}r^2 + \frac{8\sqrt{\Theta}M}{\sqrt{\pi}r^2} - \frac{4\eta\sqrt{\Theta}Q^2}{\sqrt{\pi}r^3}. \label{eq:metric_function_final}
\end{equation}
Expressing the cosmological constant in terms of the AdS radius $\ell$, the metric function becomes:
\begin{equation}
f(r) = 1 - \frac{2M}{r} + \eta \frac{Q^2}{r^2} + \frac{r^2}{\ell^2} + \frac{8\sqrt{\Theta}M}{\sqrt{\pi}r^2} - \frac{4\eta\sqrt{\Theta}Q^2}{\sqrt{\pi}r^3}. \label{eq:metric_function_ads}
\end{equation}
Let us first observe that setting $\Theta = 0$ in Eq.~\eqref{eq:metric_function_ads} recovers the metric of the phantom RN-AdS black holes studied in \cite{Hang}. 

\begin{figure}[htb!]
\begin{minipage}[t]{0.5\textwidth}
        \centering
        \includegraphics[width=\textwidth]{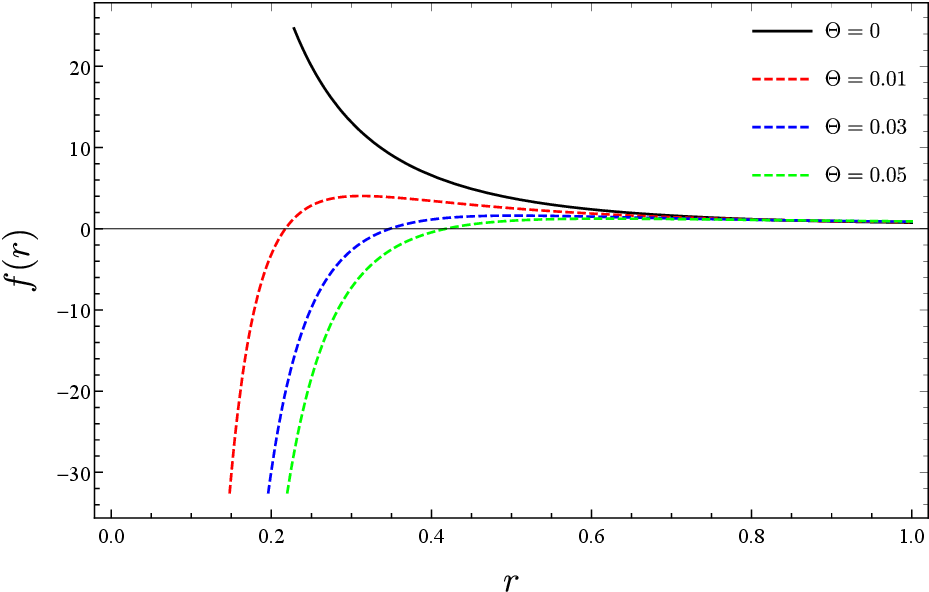}
                \subcaption{$\eta = 1$}
        \label{fig:fig1a}
\end{minipage}
\begin{minipage}[t]{0.5\textwidth}
        \centering
        \includegraphics[width=\textwidth]{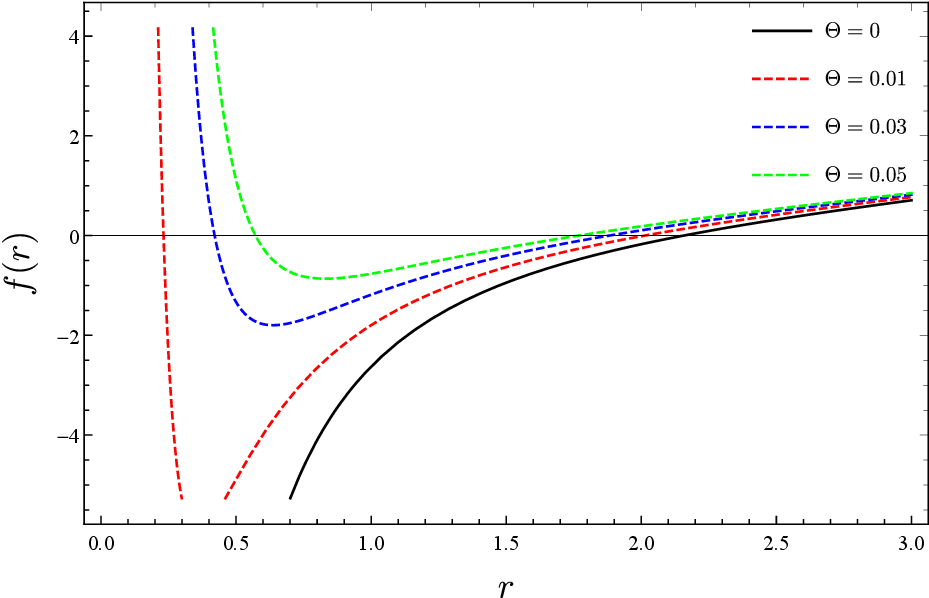}
                \subcaption{$\eta = -1$}
        \label{fig:fig1b}
\end{minipage}
\caption{Comparative plots of the metric function for different values of the noncommutative parameter with fixed parameters $\ell = 4$, $Q = 1.3$, and $M = 1$.}
\label{fig:fig1}
\end{figure}

Figure~\ref{fig:fig1} depicts the behavior of the metric function for various values of the noncommutative parameter in both the ordinary and phantom cases. 

\noindent Panel (a) depicts the ordinary case. For $\Theta = 0$, the metric function exhibits the expected behavior of an RN-AdS black hole, where no event horizon is observed for the chosen parameter set. As the noncommutative parameter $\Theta$ increases, the behavior of $f(r)$ near the origin changes significantly. The sharp divergence at $r \to 0$, associated with a classical singularity, is smoothed out by the noncommutative corrections, consistent with the idea that noncommutative geometry regularizes spacetime. For $\Theta > 0$, the metric function crosses the horizontal axis at one or more points, indicating the presence of event horizons. Notably, increasing $\Theta$ shifts the horizon positions outward, suggesting that the noncommutative effects influence the causal structure by modifying the location of the horizons. 

Panel (b) corresponds to the phantom case, where the metric function displays more pronounced modifications near the origin. For $\Theta = 0$, the function diverges negatively as $r \to 0$, consistent with the expected behavior of phantom RN-AdS black holes. However, for $\Theta > 0$, the divergence is suppressed, and the behavior of $f(r)$ near the origin becomes finite. This highlights the role of the noncommutative parameter in mitigating singularities. In the phantom scenario, the noncommutative corrections result in more exotic spacetime structures. As $\Theta$ increases, the horizons move closer together and eventually merge or disappear entirely for sufficiently large $\Theta$. This suggests a transition from black hole solutions to horizonless spacetimes, which may include traversable wormhole-like geometries often associated with phantom energy. The horizon structure in the phantom case is notably more sensitive to changes in $\Theta$ compared to the ordinary case, further emphasizing the exotic nature of these solutions.

\section{Thermodynamic Analysis} \label{sec3}

\begin{figure}[htb!]
\begin{minipage}[t]{0.5\textwidth}
        \centering
        \includegraphics[width=\textwidth]{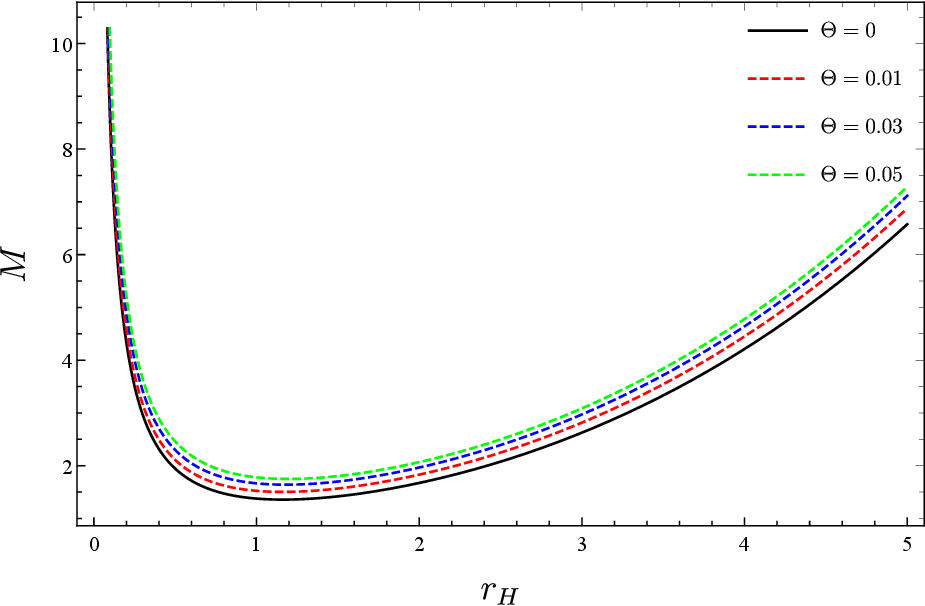}
                \subcaption{$\eta=1$}
        \label{fig:fig2a}
\end{minipage}
\begin{minipage}[t]{0.5\textwidth}
        \centering
        \includegraphics[width=\textwidth]{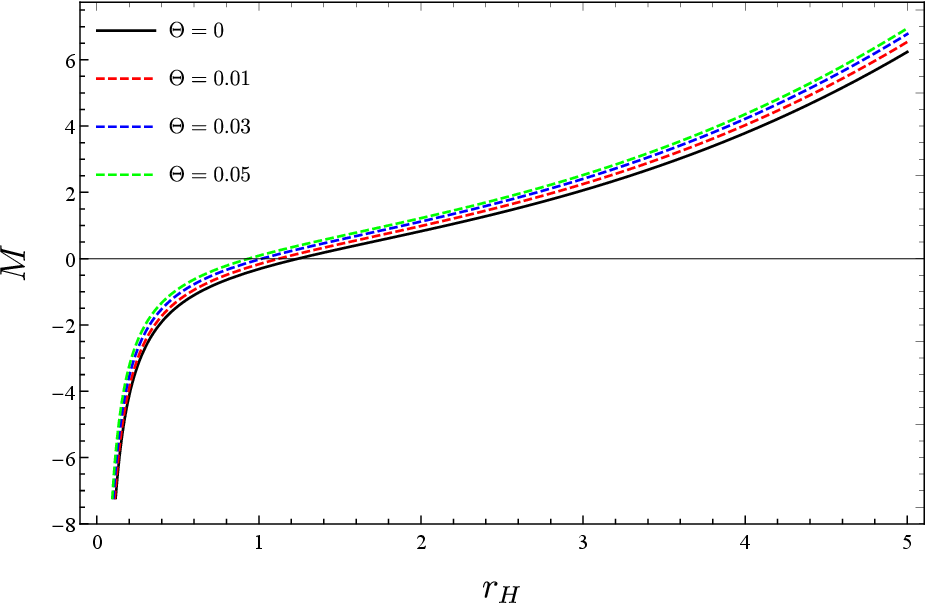}
                \subcaption{$\eta=-1$}
        \label{fig:fig2b}
\end{minipage}
\caption{Comparative plots of the mass function for different values of the noncommutative parameter with the fixed parameters $\ell=4$ and $Q=1.3$.}
\label{fig:fig2}
\end{figure}

To explore the thermodynamic properties of the black hole, we first derive the mass function by imposing the horizon condition $f(r_H) = 0$, where $r_H$ represents the event horizon radius. Solving for the mass to first-order noncommutative corrections yields the following expression:
\begin{equation}
M = \frac{r_H}{2} \left[ 1 + \frac{\eta Q^2}{r_H^2} + \frac{r_H^2}{\ell^2} + \frac{4\sqrt{\Theta} \left( \ell^2 + r_H^2 \right)}{\sqrt{\pi} \ell^2 r_H} + \frac{16 \Theta \left( \ell^2 + r_H^2 \right)}{\pi \ell^2 r_H^2} \right].
\end{equation}
Figure~\ref{fig:fig2} presents the behavior of the mass function for varying values of the noncommutative parameter $\Theta$ in both the ordinary and phantom cases in panels (a) and (b), respectively.

\noindent In the ordinary case, for $\Theta = 0$, the mass function corresponds to the classical RN-AdS black hole, exhibiting the expected non-monotonic behavior due to the interplay between the AdS curvature and the black hole charge. Specifically, the mass function initially decreases for small $r_H$ before transitioning to a monotonic increase as $r_H$ grows. When $\Theta > 0$, noncommutative corrections introduce a new feature: the initial decrease becomes more pronounced, shifting the location of the minimum to larger $r_H$. This behavior reflects the repulsive effects of noncommutative geometry at small radii, effectively reducing the black hole’s mass. At larger $r_H$, the classical monotonic growth resumes, as the noncommutative effects diminish. The interplay between noncommutativity and AdS curvature highlights the intricate modifications to black hole thermodynamics introduced by the noncommutative parameter.

Panel (b), corresponding to the phantom case, reveals a markedly different behavior. Unphysical negative mass values appear, particularly for smaller $r_H$, as a consequence of the negative kinetic energy characteristic of phantom fields.  For $\Theta = 0$, the mass function grows more slowly than in the ordinary case, consistent with the effects of phantom energy, which reduce the effective gravitational pull. As $\Theta$ increases, the noncommutative corrections lead to a stabilization effect at small $r_H$, suppressing the exotic divergences typically associated with phantom fields. Unlike the ordinary case, the mass function does not exhibit an initial decrease; instead, its growth at small radii is subdued. 

\begin{figure}[htb!]
\begin{minipage}[t]{0.5\textwidth}
        \centering
        \includegraphics[width=\textwidth]{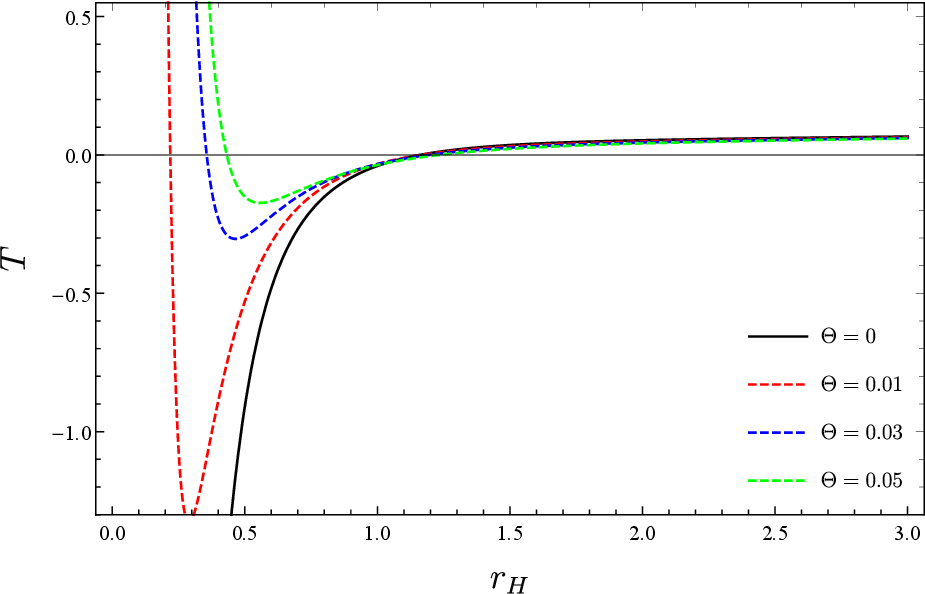}
                \subcaption{$\eta=1$}
        \label{fig:figta}
\end{minipage}
\begin{minipage}[t]{0.5\textwidth}
        \centering
        \includegraphics[width=\textwidth]{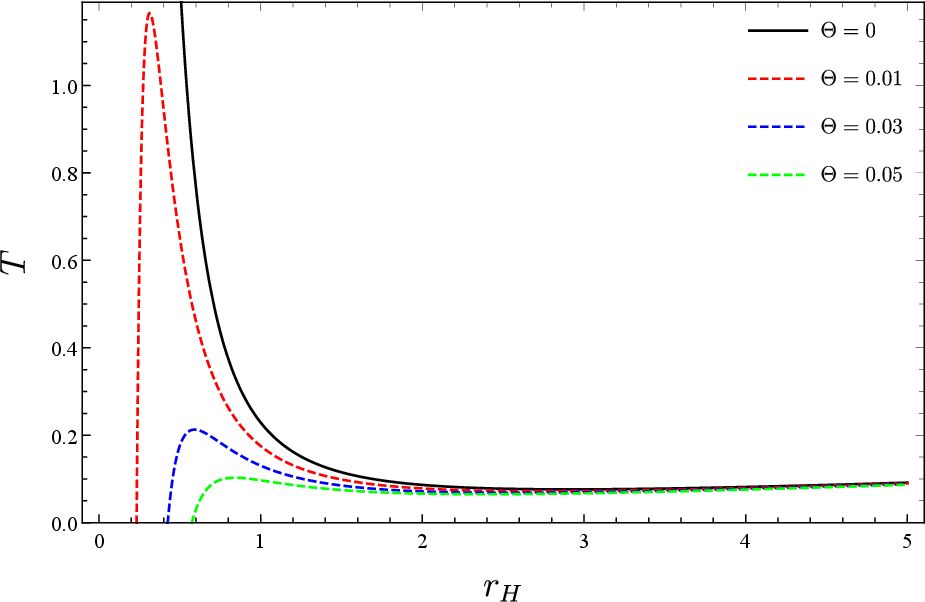}
                \subcaption{$\eta=-1$}
        \label{fig:figtb}
\end{minipage}
\caption{Comparative plots of the Hawking temperature for different values of the noncommutative parameter with the fixed parameters $\ell=4$, and $Q=1.3$.}
\label{fig:tem}
\end{figure}

To further explore the thermodynamic properties of the black hole, we calculate the Hawking temperature, which is pivotal in understanding its thermal behavior. The Hawking temperature is determined using the relation:
\begin{equation}
    T = \frac{1}{4\pi} \frac{df(r)}{dr} \bigg|_{r=r_H},
\end{equation}
where $r_H$ denotes the event horizon. By substituting the metric function and retaining terms up to the first-order noncommutative correction, the expression for the Hawking temperature becomes:
\begin{equation}
T=\frac{1}{4\pi r_{H}}\bigg(1-\eta \frac{Q^{2}}{ r_{H}^{2}}+\frac{3r_{H}^2}{\ell^{2}}\bigg) -\frac{\sqrt{\Theta }}{\pi ^{3/2}r_{H}^{2}}\bigg(1-\eta\frac{Q^{2}}{r_{H}^{2}}+\frac{r_{H}^{2}}{\ell ^{2}}\bigg)-\frac{4\Theta }{\pi ^{2}r_{H}}\bigg(\frac{1}{\ell ^{2}}+\frac{1}{r_{H}^{2}}\bigg).
\end{equation}
In Figure~\ref{fig:tem}, we compare the behavior of the Hawking temperature for different values of the noncommutative parameter $\Theta$ in both the ordinary and phantom cases, respectively.

Panel (a) of Figure~\ref{fig:tem} illustrates the Hawking temperature for the ordinary case. For $\Theta=0$, the temperature exhibits the expected behavior of a classical RN-AdS black hole. At small $r_H$, the temperature starts at negative values due to the dominance of the Coulomb-like repulsion term and transitions to positive values as $r_H$ increases, indicating the emergence of a stable thermal regime. When $\Theta > 0$, the noncommutative corrections introduce significant modifications at small $r_H$. The temperature curve shifts upward, reducing the magnitude of the negative region. This behavior can be attributed to the smearing effect of the noncommutative parameter, which regularizes the spacetime geometry and diminishes the effects of singularities. For larger $r_H$, the noncommutative effects become negligible, and the temperature asymptotically approaches the classical RN-AdS behavior. Importantly, the temperature for all $\Theta > 0$ remains higher than its classical counterpart in the positive region, suggesting that noncommutative corrections enhance the thermal radiation.

Panel (b) depicts the phantom scenario, where the influence of phantom energy significantly alters the thermal behavior. For $\Theta=0$, the Hawking temperature shows a sharp peak at small $r_H$, followed by a rapid decrease as $r_H$ increases. This reflects the destabilizing effect of phantom energy, which is associated with negative pressure and exotic thermodynamic properties. As $\Theta > 0$, the noncommutative corrections again play a stabilizing role. The peak temperature decreases with increasing $\Theta$, and the temperature becomes less sensitive to changes in $r_H$ at larger radii. The smearing effect introduced by $\Theta$ mitigates the extreme thermodynamic variations caused by the phantom field. However, for very small $r_H$, the temperature becomes unphysical (negative), highlighting the limitations imposed by the exotic nature of phantom energy. 

Next, we turn our attention to the entropy of the black hole, which incorporates the first-order noncommutative corrections and is expressed as:
\begin{equation}
S=\pi r_{H}^{2}+8\sqrt{\pi \Theta }r_{H}+32\Theta \log (r_{H}).
\end{equation}
This expression reveals several noteworthy features. Firstly, the entropy does not depend on the parameter $\eta$, indicating that the nature of the field (ordinary or phantom) has no influence on the entropy. This universality arises because the entropy is determined solely by the horizon geometry and the noncommutative corrections, which are independent of $\eta$. Secondly, in the commutative space limit, where $\Theta \to 0$, the expression reduces to the standard Bekenstein-Hawking entropy, $\mathcal{S} = \pi r_H^2$, consistent with the classical result. This recovery of the usual entropy function confirms the validity of the noncommutative corrections as a perturbative extension to the commutative case.

The additional terms involving $\Theta$ represent the effects of noncommutative geometry, introducing a linear correction proportional to $\sqrt{\Theta}$ and a logarithmic correction term. These corrections dominate at small $r_H$, reflecting the influence of spacetime noncommutativity near the black hole's event horizon. At larger $r_H$, the classical quadratic term becomes dominant, and the entropy asymptotically approaches the commutative form. 

\begin{figure}[htb!]
\begin{minipage}[t]{0.5\textwidth}
        \centering
        \includegraphics[width=\textwidth]{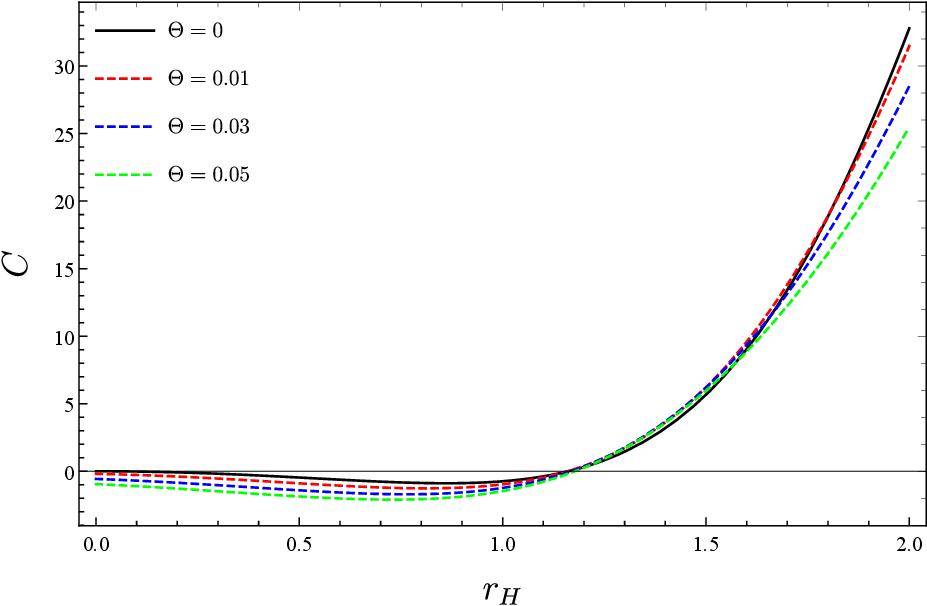}
                \subcaption{$\eta=1$}
        \label{fig:figha}
\end{minipage}
\begin{minipage}[t]{0.5\textwidth}
        \centering
        \includegraphics[width=\textwidth]{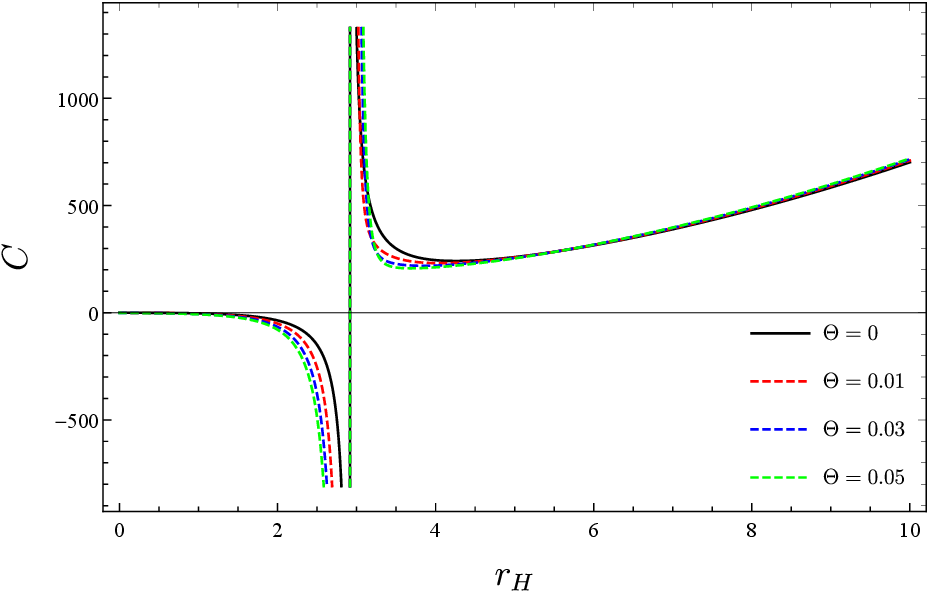}
                \subcaption{$\eta=-1$}
        \label{fig:fighb}
\end{minipage}
\caption{Heat capacity for different values of the noncommutative parameter with the fixed parameters $\ell=4$, $Q=1.3$.}
\label{fig:heat}
\end{figure}

Next, we examine the heat capacity of the black hole, which provides insights into its thermodynamic stability and phase transitions. The heat capacity, incorporating first-order noncommutative corrections, is given by:
\small
\begin{align}
&C = \frac{2\pi r_{H}^{2}\left( \ell ^{2}\left( r_{H}^{2}-\eta Q^{2}\right)
+3r_{H}^{4}\right) }{3r_{H}^{4}-\ell ^{2}\left( r_{H}^{2}-3\eta q^{2}\right) }
+\frac{16r_{H}\left( -\ell ^{4}\left( 2Q^{4}-3\eta
Q^{2}r_{H}^{2}+r_{H}^{4}\right) +\ell ^{2}\left( 9\eta
Q^{2}r_{H}^{4}-4r_{H}^{6}\right) +3r_{H}^{8}\right) }{\left( \ell ^{2}\left(
r_{H}^{2}-3\eta q^{2}\right) -3r_{H}^{4}\right) ^{2}}\sqrt{\pi \Theta } \notag \\
&+ \frac{64\left( \ell ^{6}\left( 8\eta Q^{6}-16q^{4}r_{H}^{2}+13\eta
Q^{2}r_{H}^{4}-3r_{H}^{6}\right) -3\ell ^{4}\left( 14\eta
^{2}Q^{4}r_{H}^{4}-19\eta q^{2}r_{H}^{6}+4r_{H}^{8}\right) +9\ell
^{2}r_{H}^{8}\left( 3r_{H}^{2}-2\eta Q^{2}\right) \right) }{\left( \ell
^{2}\left( r_{H}^{2}-3\eta q^{2}\right) -3r_{H}^{4}\right) ^{3}}\Theta.
\end{align}
\normalsize
The behavior of the heat capacity as a function of the event horizon radius $r_H$ is depicted in Figure~\ref{fig:heat} for both the ordinary case ($\eta=1$) and the phantom case ($\eta=-1$). 

\newpage
These plots reveal the impact of the noncommutative parameter $\Theta$ on the thermodynamic stability of the black hole. In particular, in Panel (a) of Figure~\ref{fig:heat}, the heat capacity for the ordinary case is shown. For small $r_H$, the heat capacity is negative, indicating thermodynamic instability in this regime. However, as $r_H$ increases, the heat capacity transitions to positive values, signifying a stable black hole configuration. This behavior reflects the fact that small black holes, influenced by their strong gravitational fields, are thermodynamically unstable, while larger black holes achieve stability. The noncommutative parameter $\Theta$ introduces subtle modifications to this transition. For larger values of $\Theta$, the point at which the heat capacity becomes positive shifts slightly to larger $r_H$, demonstrating that noncommutative corrections delay the onset of thermodynamic stability. At larger radii, the heat capacity increases monotonically, and the influence of $\Theta$ diminishes, converging to the classical behavior.

In Panel (b), the heat capacity for the phantom case ($\eta=-1$) exhibits a more intricate behavior. For small $r_H$, the heat capacity is negative, similar to the ordinary case, reflecting instability in this regime. However, unlike the ordinary case, the heat capacity diverges at a critical horizon radius, marking a phase transition between thermodynamic instability (negative heat capacity) and stability (positive heat capacity). Beyond this divergence, the heat capacity becomes positive and increases monotonically with $r_H$, indicating a stable black hole configuration for larger horizon radii. The noncommutative parameter $\Theta$ has a noticeable effect, slightly shifting the location of the divergence and altering the growth rate of the heat capacity in the stable region. At sufficiently large $r_H$, the heat capacity converges to its classical behavior as the effects of noncommutativity become negligible.

In continuation of our analysis, we now focus on determining the critical parameters of the black hole, specifically the critical horizon radius $r_c$, temperature $T_c$, and pressure $p_c$, which are essential for understanding the thermodynamic properties of the black hole. At first, using the relation 
\begin{equation}
\ell ^{2}=\frac{3}{8\pi p},
\end{equation}
we express the thermodynamic volume, $V$, and the pressure as%
\begin{equation}
V=\frac{4}{3\sqrt{\pi }}\mathcal{S}^{3/2}+\frac{16\sqrt{\Theta }}{3\sqrt{\pi }}\mathcal{S}+\frac{%
64\Theta }{3\sqrt{\pi }}\mathcal{S}^{1/2},
\end{equation}
\begin{equation}
p=\frac{2\Theta \left( \eta Q^{2}+16\pi r_{H}^{3}T+8r_{H}^{2}\right) }{9\pi
^{2}r_{H}^{6}}+\frac{\sqrt{\Theta }\left( -\eta Q^{2}+2\pi
r_{H}^{3}T+r_{H}^{2}\right) }{3\pi ^{3/2}r^{5}}+\frac{\eta Q^{2}+4\pi
r_{H}^{3}T-r_{H}^{2}}{8\pi r_{H}^{4}}.
\end{equation}%
We then investigate the behavior at critical points by solving for the values where certain derivatives of the pressure vanish:
\begin{equation}
\left( \frac{\partial p}{\partial r_{H}}\right) _{r_{H}=r_{c}}=0, \text{ \ \ \
and \ \ }\left( \frac{\partial ^{2}p}{\partial r_{H}^{2}}\right)
_{r_{H}=r_{c}}=0.
\end{equation}
From these conditions, we obtain expressions for the critical temperature and pressure at the critical radius:
\begin{equation}
T_{c}=-\frac{8\Theta \left( 6r_{c}^{2}-\eta q^{2}\right) }{3\left( \pi
^{2}r_{c}^{5}\right) }-\frac{2\sqrt{\Theta }\left( 5r_{c}^{2}-9\eta
Q^{2}\right) }{3\left( \pi ^{3/2}r_{c}^{4}\right) }+\frac{r^{2}-2\eta Q^{2}}{%
2\pi r_{c}^{3}},
\end{equation}%
\begin{equation}
p_{c}=-\frac{2\Theta \left( 10r_{c}^{2}-3\eta Q^{2}\right) }{3\pi
^{2}r_{c}^{6}}+\frac{\sqrt{\Theta }\left( 2\eta Q^{2}-r_{c}^{2}\right) }{\pi
^{3/2}r_{c}^{5}}+\frac{r_{c}^{2}-3\eta Q^{2}}{8\pi r_{c}^{4}},
\end{equation}%
where the critical radius is determined by ensuring that the pressure equation satisfies the following condition:
\begin{equation}
4\Theta \left( 62r_{c}^{2}-15\eta Q^{2}\right) +24\sqrt{\pi \Theta }\left(
r_{c}^{3}-3\eta Q^{2}r_{c}\right) -\frac{9\pi }{4}\left( r_{c}^{4}-6\eta
Q^{2}r_{c}^{2}\right) =0.
\end{equation}
Due to the complexity of the expressions involved, analytical solutions are not feasible, and numerical methods are required to determine the critical values of the horizon radius, temperature, and pressure. The results are summarized in Table \ref{tbbp}.

\begin{table}[tbh]
 \centering
\begin{tabular}{|l|lll|l|lll|}
\hline\hline
\rowcolor{lightgray} \multicolumn{4}{l|}{$\eta = 1$} & \multicolumn{4}{l|}{$\eta = -1$} \\ \hline
$\Theta$ & $r_{c}$ & $T_{c}$ & $p_{c}$ & $\Theta$ & $r_{c}$ & $T_{c}$ & $p_{c}$ \\ \hline\hline
0.01 & 3.41131 & 0.0289061 & 0.0015608 & 0.01 & 0.356028 & / & / \\ 
0.03 & 3.68307 & 0.0255626 & 0.00126342 & 0.03 & 0.67056 & / & / \\ 
0.05 & 3.92261 & 0.0233631 & 0.00107671 & 0.05 & 0.946589 & 0.020209 & 0.00385615 \\ 
0.07 & 4.1464 & 0.0216898 & 0.000941092 & 0.07 & 1.22839 & 0.0305717 & 0.00416909 \\ 
0.09 & 4.35879 & 0.020343 & 0.000836 & 0.09 & 1.52545 & 0.0317367 & 0.0034568 \\ 
0.10 & 4.4614 & 0.0197589 & 0.000792742 & 0.10 & 1.67903 & 0.0311617 & 0.00308621 \\  
0.11 & 4.56185 & 0.0192229 & 0.00075326 & 0.11 & 1.83462 & 0.0302726 & 0.00274946 \\ 
0.12 & 4.66028 & 0.0187285 & 0.000717534 & 0.12 & 1.99092 & 0.029238 & 0.00245329 \\ 
0.13 & 4.75683 & 0.0182704 & 0.000685045 & 0.13 & 2.1467 & 0.0281582 & 0.00219729 \\ 
\hline\hline
\end{tabular}
\caption{Critical horizon radius, temperature, and pressure for different values of the noncommutativity parameter and $\eta$.}
\label{tbbp}
\end{table}

We observe that for the ordinary case, as the value of $\Theta$ increases, the critical radius increases, while both the critical temperature and pressure decrease. This indicates that for higher values of $\Theta$, the black hole approaches a larger critical horizon, accompanied by lower thermodynamic parameters. In contrast, for the phantom case, the critical parameters exhibit a more pronounced change in $r_c$ and $T_c$, while pressure values remain undefined for certain $\Theta$ values. This suggests a more complex interaction between the parameters, particularly when $\eta = -1$, resulting in distinct thermodynamic characteristics for the black hole.

These results highlight the dependence of the critical parameters on thermodynamic variables, revealing how variations in the noncommutativity parameter influence the black hole's stability and phase transitions in both ordinary and phantom scenarios. { Before concluding this section, it is important to clarify that some of the data in Table \ref{tbbp} are presented without pressure and temperature because, for certain values of $r_{c}$ (specifically 0.356028 and 0.670560), the critical temperature and critical pressure are negative, which makes them non-physical.}

\section{Black hole as heat engine} \label{sec4}
In thermodynamics, a heat engine is a system that absorbs heat energy $Q_{H}$ from a high-temperature reservoir at temperature $T_{H}$, converts part of this energy into work $W$, and expels the remaining energy to a low-temperature reservoir at $T_{C}$. The efficiency $\Gamma $ of the engine is defined as $\Gamma =W/Q_{H}$.
\,  According to the laws of thermodynamics, the maximum efficiency achievable by any heat engine is that of a Carnot cycle, which consists of two isothermal processes (constant temperature) and two adiabatic processes (no heat transfer). The Carnot efficiency is expressed as:
\begin{equation}
\Gamma _{C}=1-\frac{T_{C}}{T_{H}}.
\end{equation}
The concept of a heat engine in the context of black hole physics was first introduced by Johnson in 2014 \cite{Johnson}. Following this pioneering work, heat engines have been investigated for various black hole configurations, including Kerr-AdS and dyonic black holes \cite{dyoni}, Gauss-Bonnet black holes \cite{Gauss}, Born-Infeld AdS black holes \cite{Infeld}, dilatonic Born-Infeld black holes \cite{dilatonic}, BTZ black holes \cite{BTZ}, and polytropic black holes \cite{polytropic}. 

\begin{figure}[H]
\centering
\includegraphics[scale=0.6]{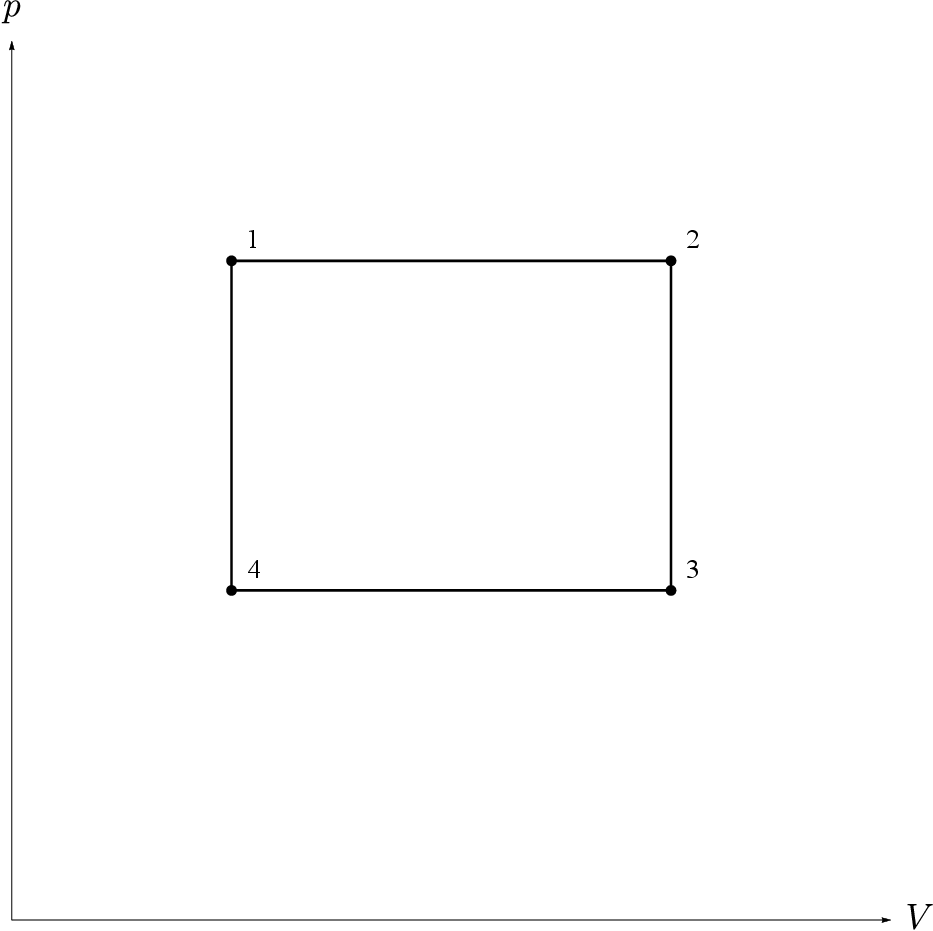}
\caption{The heat engine cycle.}
\label{fig:cycle}
\end{figure}

In this section, we explore the efficiency of a heat engine that uses noncommutative phantom RN-AdS black holes as the working substance. As illustrated in Fig.\ref{fig:cycle}, the heat engine cycle consists of two isobaric (constant pressure) and two isochoric (constant volume) processes. For simplicity, we consider a rectangular cycle ($1\rightarrow 2\rightarrow 3\rightarrow4\rightarrow 1$) in the $P-V$ plane. The work performed by the engine during one complete cycle corresponds to the area enclosed by the rectangle, which is given by:
\begin{equation}
W=\oint p \, dV.
\end{equation}%

\noindent The total work done over the cycle is expressed as:
\begin{equation}
W=W_{1\rightarrow 2}+W_{3\rightarrow 4}=p_{1}\left( V_{2}-V_{1}\right)
+p_{4}\left( V_{4}-V_{3}\right) ,
\end{equation}
which can be written in terms of the standard Bekenstein-Hawking entropy as:
\begin{equation}
W=\frac{4}{3\sqrt{\pi }}\left( p_{1}-p_{4}\right) \left(\mathcal{S}_{2}^{3/2}-\mathcal{S}_{1}^{3/2}\right) +\frac{16\sqrt{\Theta }}{3\sqrt{\pi }}\left(p_{1}-p_{4}\right) \left( \mathcal{S}_{2}-\mathcal{S}_{1}\right) +\frac{64\Theta }{3\sqrt{\pi }}\left( p_{1}-p_{4}\right) \left( \mathcal{S}_{2}^{1/2}-\mathcal{S}_{1}^{1/2}\right) .
\end{equation}
Given the heat capacity at constant volume, $C_{V}=0$, no heat exchange occurs during the isochoric processes. Consequently, the heat absorbed by the system is entirely attributed to the isobaric processes, which can be expressed with:
\begin{equation}
Q_{H}=\int\limits_{T_{1}}^{T_{2}}C_{p}dT=\int\limits_{S_{1}}^{S_{2}}C_{p}%
\left( \frac{dT}{dS}\right) dS=\int\limits_{S_{1}}^{S_{2}}TdS=M_{2}-M_{1},
\end{equation}%
that appears in the explicit expression as: 
\begin{eqnarray}
Q_{H} &=&\frac{\sqrt{\mathcal{S}_{2}}-\sqrt{\mathcal{S}_{1}}}{\sqrt{\pi }}+\eta Q^{2}\left(\sqrt{\frac{\pi }{\mathcal{S}_{2}}}-\sqrt{\frac{\pi }{\mathcal{S}_{1}}}\right) +\frac{8}{3\sqrt{\pi }}p_{1}\left( \mathcal{S}_{2}^{3/2}-\mathcal{S}_{1}^{3/2}\right)   \notag \\
&&+\frac{32}{3}\sqrt{\frac{\Theta }{\pi }}p_{1}\left( \mathcal{S}_{2}-\mathcal{S}_{1}\right) + \frac{16\Theta }{\sqrt{\pi }}\left[ \left( \frac{1}{\sqrt{\mathcal{S}_{2}}}-\frac{1}{\sqrt{S_{1}}}\right) +\frac{8}{3}p_{1}\left( \sqrt{\mathcal{S}_{2}}-\sqrt{\mathcal{S}_{1}}\right) \right] .
\end{eqnarray}%
Thus, the efficiency of the heat engine is then given by:
\begin{equation}
\Gamma =\frac{W}{Q_{H}}=\frac{\left( p_{1}-p_{4}\right) \left(
V_{2}-V_{1}\right) }{M_{2}-M_{1}}.  \label{effici}
\end{equation}%
This efficiency can be compared with the Carnot efficiency, $\Gamma _{C}$,  by setting the higher temperature $T_{H}$ as  $T_{2}$, and the lower temperature $T_{C}$ as $T_{4}$. The Carnot efficiency reads:
\begin{equation}
\Gamma _{C}=1-\frac{T_{4}\left( p_{4},\mathcal{S}_{1}\right) }{T_{2}\left(
p_{1},\mathcal{S}_{2}\right) }.  \label{38}
\end{equation}
Since $T_{4}<T_{2}$, it follows that $0<\Gamma _{C}<1$. 

\begin{figure}[H]
\begin{minipage}[t]{0.5\textwidth}
        \centering
        \includegraphics[width=\textwidth]{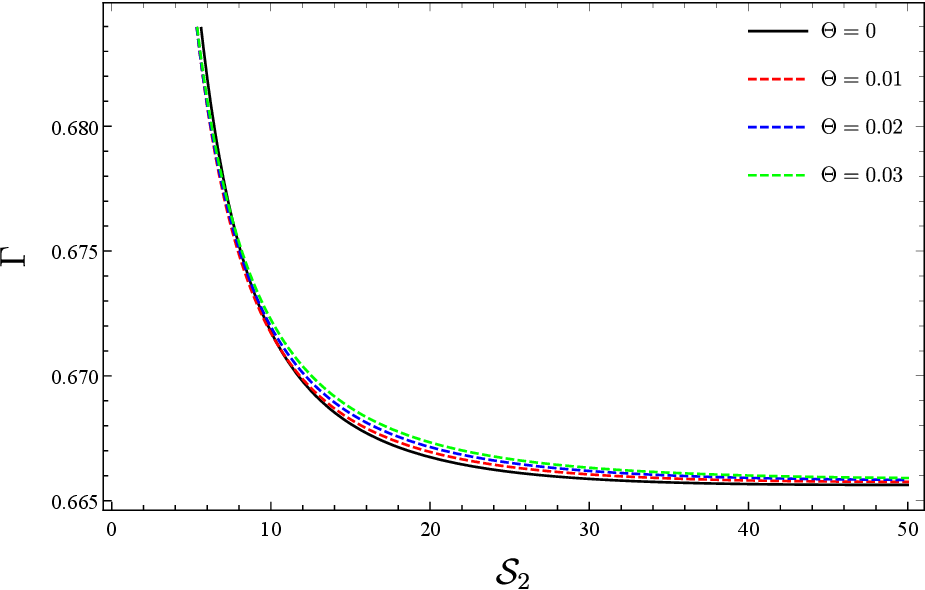}
                \subcaption{$\eta=1$}
        \label{fig:gama1}
\end{minipage}
\begin{minipage}[t]{0.5\textwidth}
        \centering
        \includegraphics[width=\textwidth]{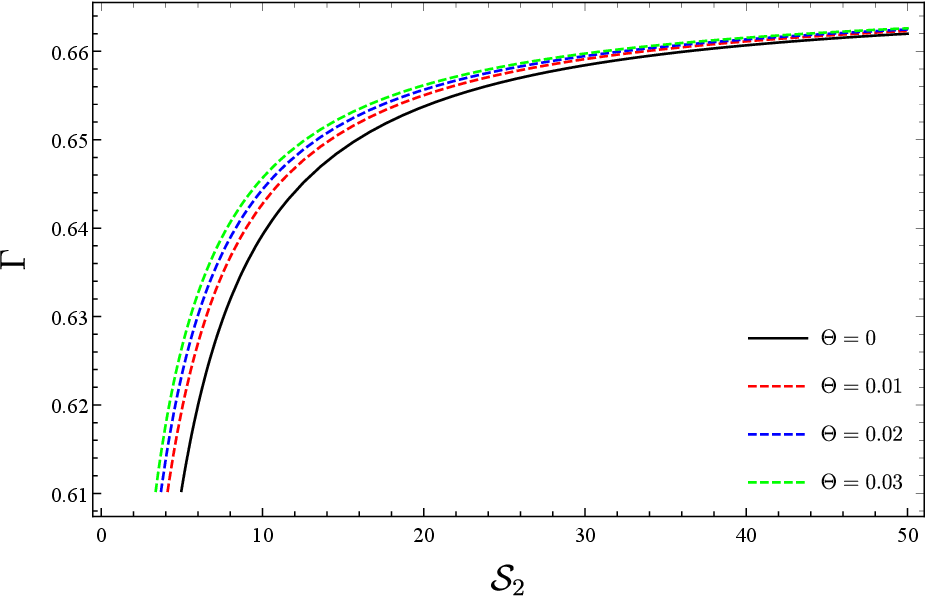}
                \subcaption{$\eta=-1$}
        \label{fig:gamm2}
\end{minipage}
\caption{Variation of the heat engine efficiency with the noncommutative parameter.}
\label{fig:gamma}
\end{figure}

\begin{figure}[H]
\begin{minipage}[t]{0.5\textwidth}
        \centering
        \includegraphics[width=\textwidth]{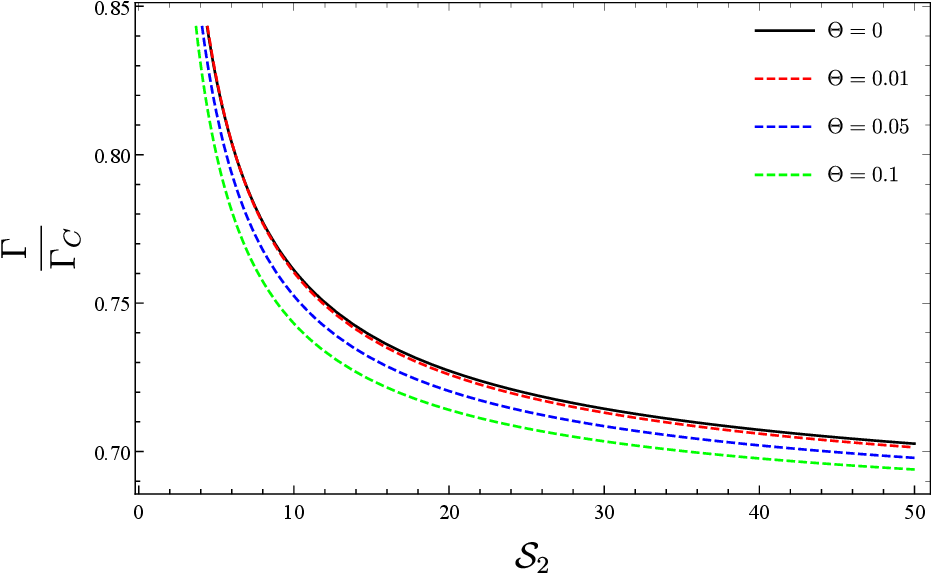}
                \subcaption{$\eta=1$}
        \label{fig:rat1}
\end{minipage}
\begin{minipage}[t]{0.5\textwidth}
        \centering
        \includegraphics[width=\textwidth]{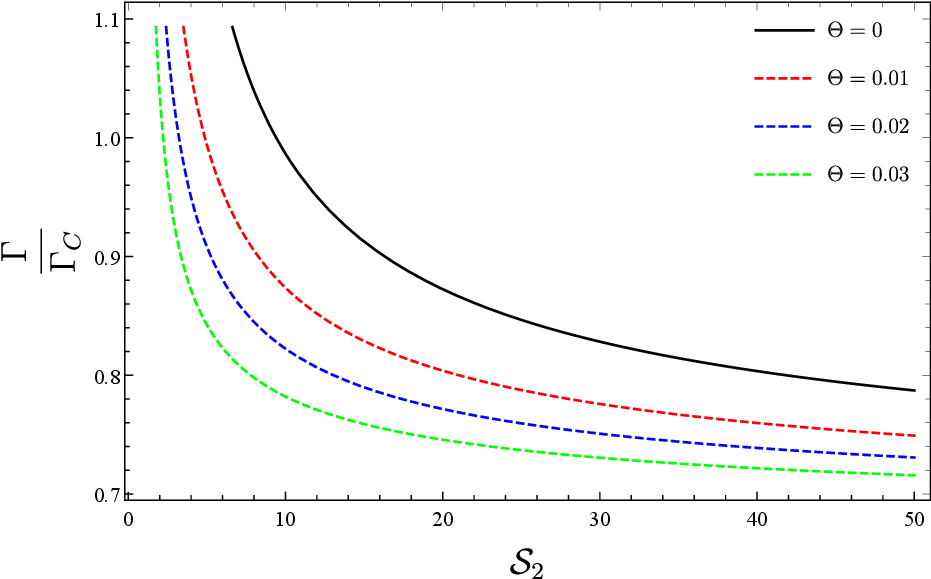}
                \subcaption{$\eta=-1$}
        \label{fig:rat2}
\end{minipage}
\caption{Ratio of the heat engine efficiency to the Carnot efficiency as a function of the noncommutative parameter.}
\label{fig:rat}
\end{figure}

With all the necessary quantities for our analysis derived, we now examine how the phantom/Maxwell efficiency varies with different noncommutativity parameters. To this end, Figure \ref{fig:gamma} illustrates the heat engine efficiency $\Gamma $, while Figure \ref{fig:rat} depicts the ratio $\frac{\Gamma }{\Gamma_{C}}$, respectively. 

\noindent Panel (a) of Figure \ref{fig:gamma} illustrates the variation of the heat engine efficiency with the noncommutative parameter in the ordinary case. The efficiency exhibits a steady increase as $\Theta$ grows, reflecting the influence of noncommutative geometry, which enhances the energy extraction capabilities of the black hole heat engine. 

Panel (b) shows the efficiency $\Gamma$  for the phantom case, where a similar increasing trend is observed. However, the overall efficiency remains lower compared to the ordinary case due to the exotic nature of phantom fields, characterized by negative energy density effects. Despite this, noncommutative corrections provide a stabilizing influence, leading to an overall enhancement of the efficiency with increasing $\Theta$. 

\noindent Panel (a) of  Figure \ref{fig:rat} depicts the ratio $\Gamma / \Gamma_C $ as a function of the noncommutative parameter. The ratio decreases slightly as $ \Theta $ increases, suggesting that while noncommutative effects enhance the absolute efficiency $ \Gamma $, the relative performance of the heat engine compared to the Carnot cycle diminishes. This trend indicates that the improvements in $ \Gamma $ due to noncommutativity are less pronounced relative to the theoretical maximum efficiency.

Panel (b) presents the ratio characteristics for the phantom case. Here, the ratio also decreases with increasing $ \Theta $, reflecting the combined influence of phantom fields and noncommutative geometry. Phantom fields inherently suppress the engine's efficiency, and while noncommutative effects improve the absolute efficiency $ \Gamma $, they do not fully counterbalance the destabilizing effects of the phantom fields. As a result, the relative performance of the heat engine compared to the Carnot efficiency declines with increasing $ \Theta $.
These behaviors underscore the role of noncommutativity in improving thermodynamic performance.

\section{Equations of the Orbital Motion} \label{sec5}

Next, within this section, we analyze timelike and null geodesics to explore the motion of particles in the spacetime described by the metric in Eq.~\eqref{eq:metric_function_ads}. The geodesic equations governing particle trajectories around phantom RN-AdS black holes in a noncommutative space are given by:expressed as:
\begin{equation}
\ddot{t} + \frac{f^{\prime}(r)}{f(r)} \dot{r} \dot{t} = 0, \label{time}
\end{equation}
\begin{equation}
\ddot{r} + f(r) \left( \frac{f^{\prime}(r) \dot{t}^2 + \frac{\dot{r}}{f^{\prime}(r)} - 2r\dot{\theta}^2 - 2r\sin^2\theta \dot{\varphi}^2}{2} \right) = 0, \label{radial}
\end{equation}
\begin{equation}
\ddot{\theta} + \frac{2}{r} \dot{r} \dot{\theta} - \cos\theta \sin\theta \dot{\varphi}^2 = 0, \label{teta}
\end{equation}
\begin{equation}
\ddot{\varphi} + \frac{2}{r} \dot{r} \dot{\varphi} + 2\cot\theta \dot{\theta} \dot{\varphi} = 0. \label{phiii}
\end{equation}
In this context the prime symbol $(')$ denotes differentiation with respect to the radial coordinate $r$. Due to the symmetric nature of the spherical geometry, the solutions can be restricted to an equatorial plane passing through a great circle of the sphere. This symmetry allows us to select initial conditions such that $\theta = \pi /2$ and $\dot{\theta}(0) = 0$, simplifying the geodesic Eqs. (\ref{time}) and (\ref{phiii}) to the following forms:
\begin{equation}
\dot{t} = \frac{E}{f(r)},
\end{equation}
\begin{equation}
\dot{\varphi} = \frac{L}{r^2}.
\end{equation}
Here, $E$ and $L$ are integration constants associated with the conserved quantities of total energy and angular momentum for a test particle. Considering these equations, along with the constraint for null and timelike geodesics, $g_{\mu \nu} \dot{x}^\mu \dot{x}^\nu = -\delta$, the expression for the radial velocity $\dot{r}$ is given by:
\begin{equation}
\dot{r}^2 = E^2 - \left( 1 - \frac{2M}{r} + \eta \frac{Q^2}{r^2} + \frac{r^2}{\ell^2} + \frac{8\sqrt{\Theta}M}{\sqrt{\pi}r^2} - \frac{4\eta \sqrt{\Theta}Q^2}{\sqrt{\pi}r^3} \right) \left( \delta + \frac{L^2}{r^2} \right), \label{rad1}
\end{equation}
where
\begin{equation}
\delta = 
\begin{cases} 
1, & \text{for timelike particles}, \\ 0, & \text{for lightlike particles}.
\end{cases}
\end{equation}
From Eq.~\eqref{rad1}, the effective potential can be expressed as:
\begin{equation}
V_{\mathrm{eff}}(r) = \left( 1 - \frac{2M}{r} + \eta \frac{Q^2}{r^2} + \frac{r^2}{\ell^2} + \frac{8\sqrt{\Theta}M}{\sqrt{\pi}r^2} - \frac{4\eta \sqrt{\Theta}Q^2}{\sqrt{\pi}r^3} \right) \left( \delta + \frac{L^2}{r^2} \right). \label{pot1}
\end{equation}
Eqs.~\eqref{rad1} and \eqref{pot1} describe the motion of a classical particle with energy $E$ in a one-dimensional potential $V_{\mathrm{eff}}(r)$. The effective potential provides key insights into the dynamics of the test particle and the influence of the black hole's parameters, including noncommutativity.

\subsection{Orbits of the timelike particle}

\begin{figure}[htb!]
\begin{minipage}[t]{0.5\textwidth}
        \centering
        \includegraphics[width=\textwidth]{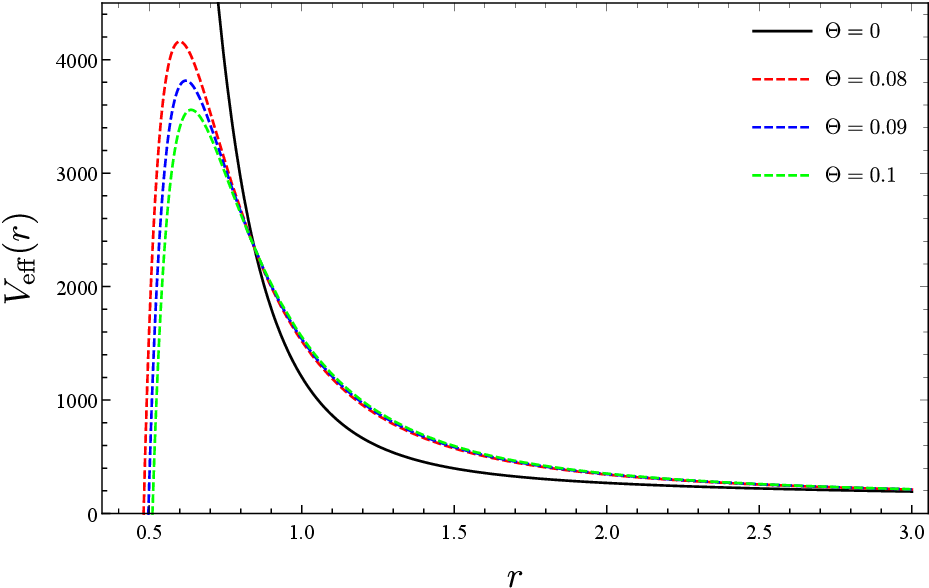}
                \subcaption{$\eta=1$}
        \label{fig:va}
\end{minipage}
\begin{minipage}[t]{0.5\textwidth}
        \centering
        \includegraphics[width=\textwidth]{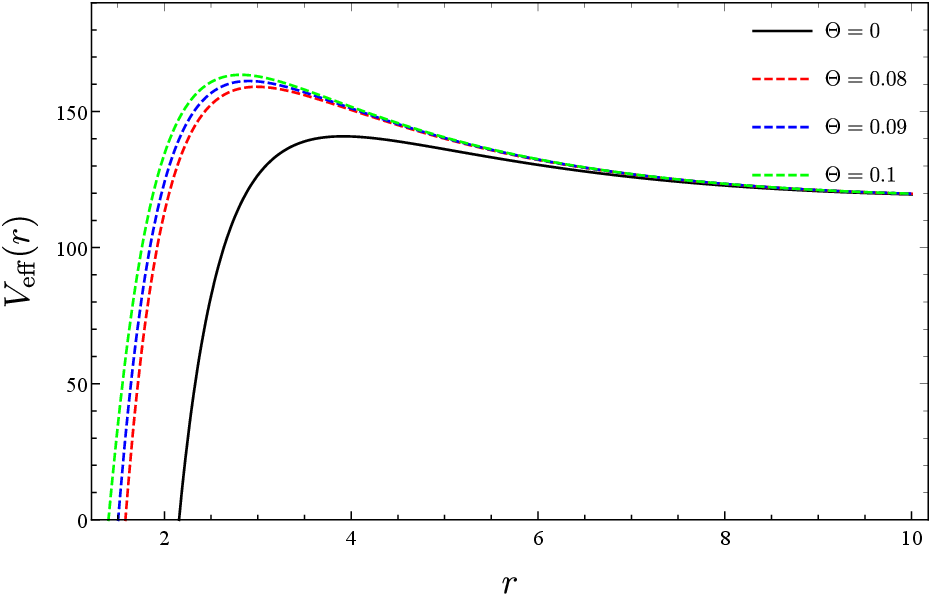}
                \subcaption{$\eta=-1$}
        \label{fig:vb}
\end{minipage}\\
\begin{minipage}[t]{0.5\textwidth}
        \centering
        \includegraphics[width=\textwidth]{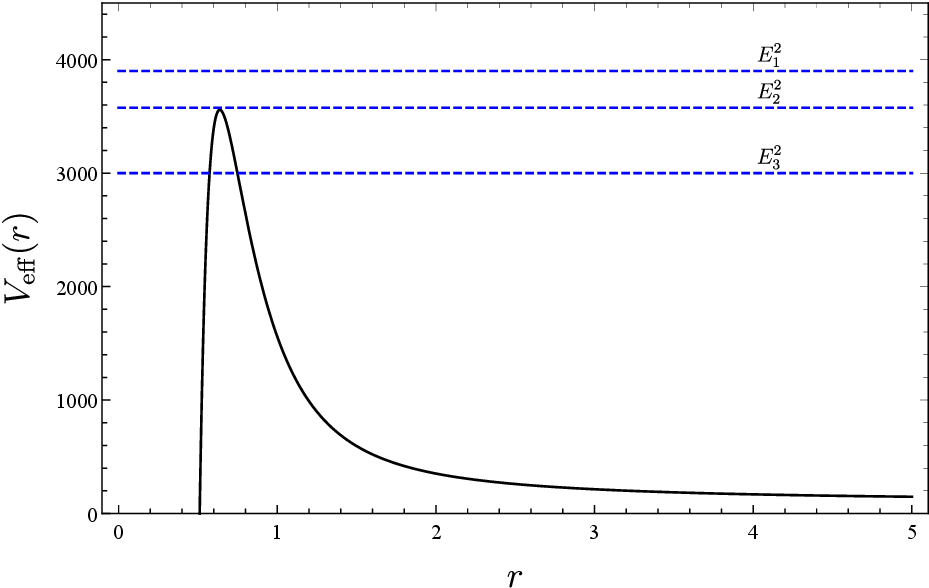}
                \subcaption{$\eta=1$ and $\Theta=0.1$}
        \label{fig:vc}
\end{minipage}
\begin{minipage}[t]{0.5\textwidth}
        \centering
        \includegraphics[width=\textwidth]{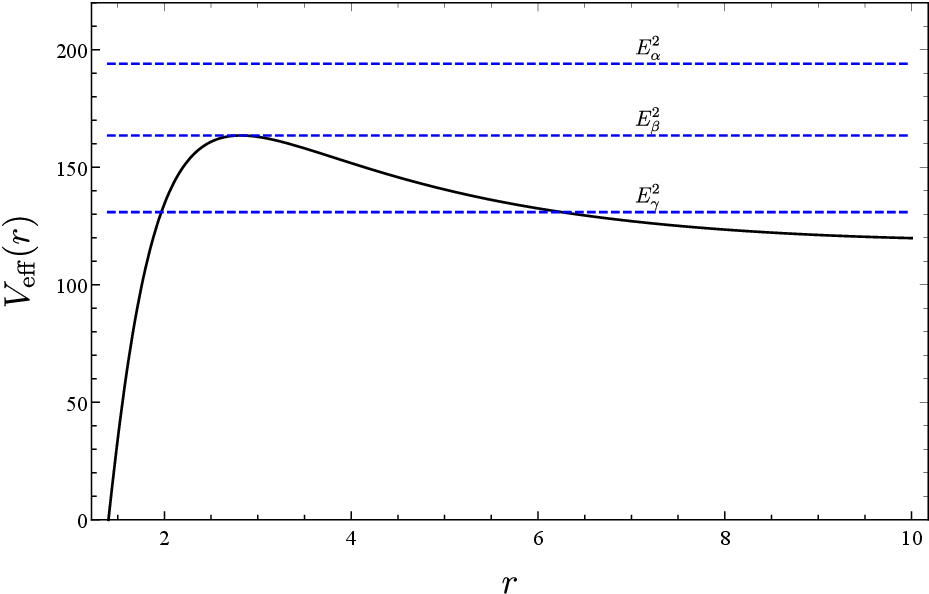}
                \subcaption{$\eta=-1$ and $\Theta=0.1$}
        \label{fig:vd}
\end{minipage} 
\caption{{Plots of the timelike effective potential $ V_{\text{eff}}(r) $ as a function of the radial coordinate $ r $ for different values of $ \eta $ and $ \Theta $. The horizontal dashed lines represent the squared energy levels $ E^2 $ of test particles in the potential.}}
\label{fig:Veff}
\end{figure}

For timelike particles, where $\delta =1$, the effective potential for timelike geodesics can be derived directly from Eq.~\eqref{pot1} as:
\begin{equation}
V_{\mathrm{eff}}\left( r\right) =\left( 1-\frac{2M}{r}+\eta \frac{Q^{2}}{r^{2}}+\frac{r^{2}}{l^{2}}+\frac{8\sqrt{\Theta }M}{\sqrt{\pi }r^{2}}-\frac{4\eta \sqrt{\Theta }Q^{2}}{\sqrt{\pi }r^{3}}\right) \left( 1+\frac{L^{2}}{r^{2}}\right) .
\end{equation}
The timelike effective potential is influenced by various parameters, including mass, charge, cosmological constant (AdS radius), angular momentum, and noncommutativity. To clarify the impact of $\Theta$, we fix the other parameters as $M=1$, $Q=1.3$, $l=4$, and $L=40$, and present the plots of the effective potential in Figure~\ref{fig:Veff} as a function of the radial coordinate for different values of noncommutativity parameter in the ordinary and phantom scenarios.

\newpage
\noindent The top row of Figure~\ref{fig:Veff}, namely panels (a) and (b), shows how the effective potential varies with $ \Theta $. In panel (a), where \( \eta = 1 \), increasing \( \Theta \) leads to a decrease in the peak of the effective potential. This suggests that a larger noncommutativity parameter \( \Theta \) reduces the potential barrier, making it easier for particles to escape. Conversely, in panel (b), where \( \eta = -1 \), increasing \( \Theta \) has the opposite effect, raising the peak of \( V_{\mathrm{eff}} \). This indicates that in the phantom field case, higher values of \( \Theta \) enhance the potential barrier, thereby making escape more difficult and possibly leading to more stable bound orbits.


The bottom row of Figure~\ref{fig:Veff}, comprising panels (c) and (d), illustrates specific cases with $ \Theta = 0.1$, offering deeper insight into orbital dynamics. In panel (c), for $ \eta = 1 $, the effective potential features a well-defined peak, with horizontal dashed lines denoting energy levels $ E_1^2, E_2^2, E_3^2 $. These thresholds determine the particle’s fate: at $ E = E_1 $, the motion leads to absorbing orbits; at $ E = E_2 $, the condition $ E_2^2 = V_{\mathrm{eff}}(r) $ corresponds to unstable circular orbits; and at $ E = E_3 $, the particle follows an escape trajectory. In panel (d), for $\eta = -1$, the potential barrier is lower but still structured to permit energy-dependent orbital transitions. The presence of distinct energy levels in both cases underscores the influence of parameter choices on geodesic motion.

Overall, this figure effectively illustrates how the $ \Theta $ and $ \eta $ impact the geodesic structure in a noncommutative phantom RN-AdS black hole spacetime. The results suggest that for standard fields $\eta = 1$, increasing $ \Theta $ facilitates particle escape, while for phantom fields $\eta = -1$, a higher $ \Theta $ enhances stability by increasing the potential barrier. These findings provide valuable insights into the influence of noncommutativity on black hole dynamics and the nature of particle trajectories in such spacetimes.

\begin{figure}[htb!]
\begin{minipage}[t]{0.5\textwidth}
        \centering
        \includegraphics[width=\textwidth]{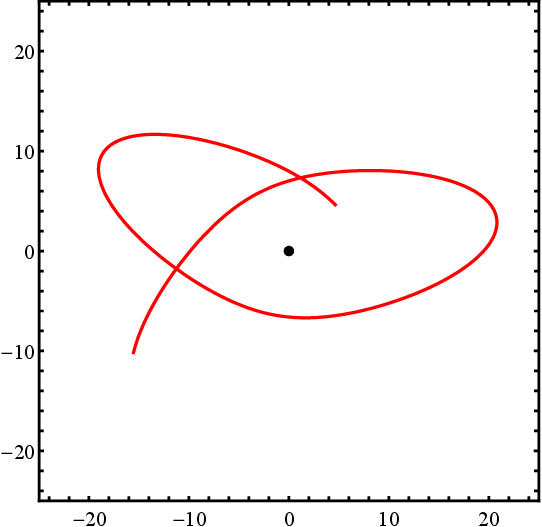}
                \subcaption{$\eta=1$}
        \label{fig:ta}
\end{minipage}
\begin{minipage}[t]{0.5\textwidth}
        \centering
        \includegraphics[width=\textwidth]{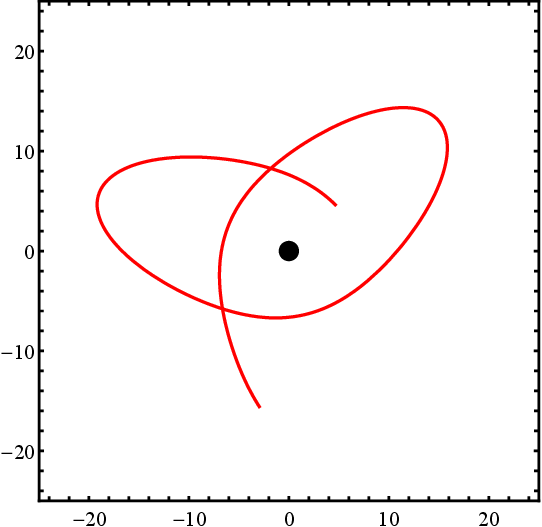}
                \subcaption{$\eta=-1$}
        \label{fig:tb}
\end{minipage}
\caption{{Unstable circular orbits of test particles in the $(x,y)$ plane. The axes represent the spatial coordinates in the equatorial plane, with the black dot indicating the black hole's position.}}
\label{fig:traj1}
\end{figure}

Figure~\ref{fig:traj1} explicitly presents the unstable circular orbits of test particles corresponding to energy $E_2$ for $\eta = 1$ and $E_{\delta}$ for $\eta = -1$.

\noindent 
In both cases, we observe that small perturbations can cause the particle to deviate significantly from the circular trajectory, leading either to escape or eventual absorption. The red trajectories depict how a slight deviation from the equilibrium orbit results in chaotic motion, emphasizing the instability of these configurations. The lower potential barrier for $ \eta = -1 $ suggests that perturbations more easily drive the particle away from the circular orbit compared to the $ \eta = 1 $ case. This highlights the role of the noncommutativity parameter in shaping the orbital dynamics and stability of geodesic motion.

Next, we analyze the orbital motion and its stability and investigate the impact of noncommutativity on particle trajectories. We begin by considering circular motion $(\dot{r} = 0)$, where the effective potential satisfies the following conditions:
\begin{equation}
V_{\mathrm{eff}}\left( r\right) =E^{2}.
\end{equation}%
Additionally, for a stable circular orbit, the radial derivative of the effective potential must vanish:
\begin{equation}
\frac{d}{dr}V_{\mathrm{eff}}\left( r\right) =0.
\end{equation}
Taking these conditions into account, the specific energy and angular momentum for test particles in circular orbits are obtained as:
\begin{equation}
L^{2}=\frac{r^{3}f^{\prime }\left( r\right) }{2f\left( r\right) -rf^{\prime
}\left( r\right) },  \label{23}
\end{equation}%
\begin{equation}
E^{2}=\frac{2f\left( r\right) ^{2}}{2f\left( r\right) -rf^{\prime }\left(
r\right) }.  \label{24}
\end{equation}
However, it's of great importance to identify the ISCO, where the orbit's maximum and minimum effective potential intersect \cite{Javed2025}. To determine the ISCO radius of the particle, an additional condition is necessary, which is provided by the second derivative of $V_{\mathrm{eff}}\left( r\right) $ with respect to the radial coordinate $r$,
\begin{equation}
\frac{d^{2}}{dr^{2}}V_{\mathrm{eff}}\left( r\right) =0.  \label{25}
\end{equation}%
By solving Eq.~\eqref{25} in conjunction with Eqs.~\eqref{23} and \eqref{24}, the following condition is found:
\begin{equation}
\left. rf\left( r\right) f^{\prime \prime }\left( r\right) +3f^{\prime
}\left( r\right) f\left( r\right) -2rf^{\prime }\left( r\right)
^{2}\right\vert _{r=r_{\mathrm{ISCO}}}=0.
\end{equation}%
The ISCO parameters $r_{\mathrm{ISCO}}$, $E_{\mathrm{ISCO}}$ and $L_{\mathrm{ISCO}}$ are determined numerically and presented in Table \ref{tablisco} for the ordinary and phantom scenarios.
\begin{table}[tbh]
\centering
\begin{tabular}{|l|lll|l|lll|}
\hline\hline
\rowcolor{lightgray} \multicolumn{4}{l|}{$\eta =1$ and $Q=1.3$}& \multicolumn{4}{l|}{$\eta =-1$ and $Q=1.3$} \\ \hline
$\Theta $ & $r_{\mathrm{ISCO}}$ & $L_{\mathrm{ISCO}}$ & $E_{\mathrm{ISCO}}$&$\Theta $ & $r_{\mathrm{ISCO}}$ & $L_{\mathrm{ISCO}}$ & $E_{\mathrm{ISCO}}$
\\ \hline\hline
0.14 & 1.01952 & 0.0933595 & 0.0179358&0.10 & 3.96248 & 18.0218 & 4.69971 \\ 
0.15 & 1.03000 & 0.143387 & 0.0410391&0.11 & 3.87895 & 17.4634 & 4.65773 \\ 
0.16 & 1.03931 & 0.177407 & 0.0611152&0.12 & 3.79344 & 16.9087 & 4.61740 \\ 
0.17 & 1.04847 & 0.203779 & 0.0785487&0.13 & 3.70492 & 16.3532 & 4.57887 \\ 
0.18 & 1.05719 & 0.225368 & 0.0937654&0.14 & 3.61211 & 15.7921 & 4.54250 \\ 
0.20 & 1.07343 & 0.259234 & 0.118774&0.15 & 3.51317 & 15.2195 & 4.50912 \\
0.22 & 1.08825 & 0.284883 & 0.138105& 0.16 & 3.40531 & 14.6275 & 4.48032 \\ 
0.24 & 1.10183 & 0.305017 & 0.153143&0.17 & 3.28369 & 14.0042 & 4.45960 \\ 
0.26 & 1.11436 & 0.321165 & 0.164864& 0.18 & 3.13790 & 13.3282 & 4.45642 \\ 
\hline\hline
\end{tabular}
\caption{Numerical values of the ISCO parameters $r_{\rm{ISCO}}$, $L_{\rm{ISCO}}$ and $E_{\rm{ISCO}}$ for test particles under various scenarios.}
\label{tablisco} 
\end{table}

For the ordinary case, the ISCO parameters increase as $\Theta$ grows. Both the ISCO radius and angular momentum increase, indicating that test particles move to orbits farther from the black hole, while the energy also increases, reflecting the greater energy required for stable orbits at larger $\Theta$. In the phantom scenario, the ISCO parameters decrease as $\Theta$ increases. Both the ISCO radius and angular momentum decrease, implying that the gravitational environment becomes more restrictive, pulling test particles closer to the black hole. This results in lower energy and angular momentum for the particles compared to the ordinary case. The contrast between the two cases highlights the influence of the $\eta$ parameter on the stability and dynamics of orbits around the black hole.

\subsection{ Black hole shadow}
\begin{figure}[htbp]
\begin{minipage}[t]{0.5\textwidth}
        \centering
        \includegraphics[width=\textwidth]{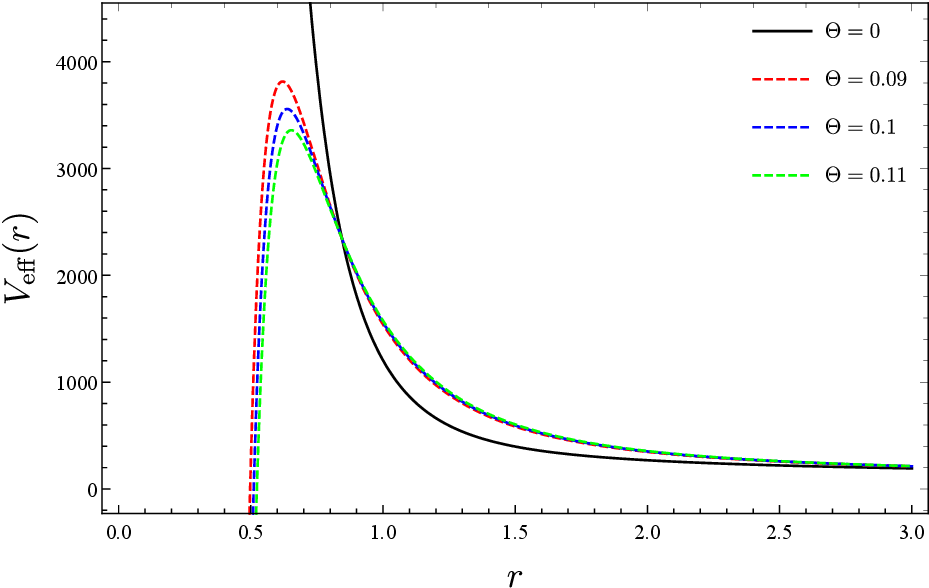}
                \subcaption{$\eta=1$}
        \label{fig:vaa}
\end{minipage}
\begin{minipage}[t]{0.5\textwidth}
        \centering
        \includegraphics[width=\textwidth]{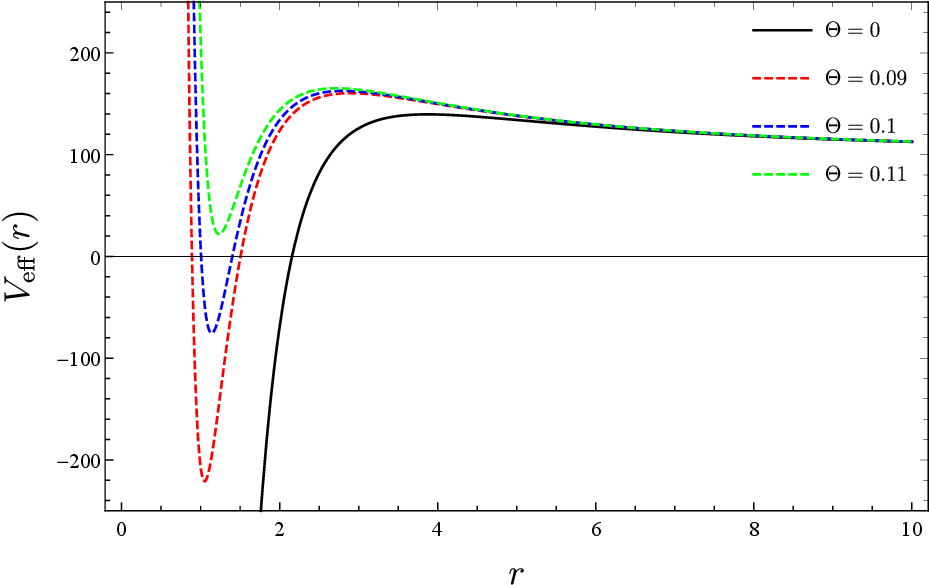}
                \subcaption{$\eta=-1$}
        \label{fig:vbb}
\end{minipage}
\caption{Variation of $V_{\mathrm{eff}}(r)$ with respect to the radial distance 
$r$ for different values of the parameter $\Theta $.  The fixed parameters are $M=1$, $Q=1.3$, $\ell=4$ and $L=40$.}
\label{fig:Veffp}
\end{figure}

The effective potential for lightlike particles, governing the null geodesics, is expressed as
\begin{equation}
V_{\mathrm{eff}}\left( r\right) =\left( 1-\frac{2M}{r}+\eta \frac{Q^{2}}{%
r^{2}}+\frac{r^{2}}{\ell^{2}}+\frac{8\sqrt{\Theta }M}{\sqrt{\pi }r^{2}}-\frac{4\eta \sqrt{\Theta }Q^{2}}{\sqrt{\pi }r^{3}}\right) \frac{L^{2}}{r^{2}}.
\end{equation}
This potential characterizes the radial motion of photons within the black hole's spacetime. Analyzing its behavior provides critical insights into the photon sphere's properties in the specified black hole geometry. Figure \ref{fig:Veffp} illustrates the variation of the effective potential $V_{\mathrm{eff}}(r)$ as a function of the radial coordinate $r$ for different values of the parameter $\Theta $. In the phantom scenario, an increase in the noncommutativity parameter leads to a reduction in the peak value of the effective potential. Despite this decrease, the parameter $\Theta $ significantly influences both the potential and the photon sphere of the black hole. The effective potential exhibits one maximum and one minimum, corresponding to unstable and stable circular orbits, respectively. Notably, as the value of $\Theta $ increases, the maximum value of the potential increases substantially.

Next, we proceed with analyzing the black hole's shadow. The shadow's characteristics are determined by the turning point of photon trajectories, represented by $r = r_{\mathrm{ph}}$, corresponding to the photon sphere radius. At this turning point, the following conditions must be satisfied:  
\begin{equation}  
\left. V_{\mathrm{eff}}\left( r\right) \right|_{r=r_{\mathrm{ph}}} = E^{2}, \quad \text{and} \quad \left. V_{\mathrm{eff}}^{\prime }\left( r\right) \right|_{r=r_{\mathrm{ph}}} = 0,  
\end{equation}  
ensuring that $ V_{\mathrm{eff}}\left( r \right) $ attains a maximum at $ r = r_{\mathrm{ph}} $, i.e.,  
\begin{equation}  
\left. V_{\mathrm{eff}}^{\prime \prime }\left( r\right) \right|_{r=r_{\mathrm{ph}}} < 0.  
\end{equation}  
The impact parameter, defined as $ b_{\mathrm{cr}} = \frac{L}{E} $, is obtained from the first condition as  
\begin{equation}  
b_{\mathrm{cr}} = \frac{r_{\mathrm{ph}}}{\sqrt{1-\frac{2M}{r_{\mathrm{ph}}} + \eta \frac{Q^{2}}{r_{\mathrm{ph}}^{2}} + \frac{r_{\mathrm{ph}}^{2}}{\ell^{2}}}}  
- 2\sqrt{\frac{\Theta}{\pi}} \frac{\frac{2M}{r_{\mathrm{ph}}} - \eta \frac{Q^{2}}{r_{\mathrm{ph}}^{2}}}{\left(1 - \frac{2M}{r_{\mathrm{ph}}} + \eta \frac{Q^{2}}{r_{\mathrm{ph}}^{2}} + \frac{r_{\mathrm{ph}}^{2}}{\ell^{2}}\right)^{3/2}}  
+ \frac{6\Theta}{\pi r_{\mathrm{ph}}} \frac{\left(\frac{2M}{r_{\mathrm{ph}}} - \eta \frac{Q^{2}}{r_{\mathrm{ph}}^{2}}\right)^{2}}{\left(1 - \frac{2M}{r_{\mathrm{ph}}} + \eta \frac{Q^{2}}{r_{\mathrm{ph}}^{2}} + \frac{r_{\mathrm{ph}}^{2}}{\ell^{2}}\right)^{5/2}}. \label{48}  
\end{equation}  
The boundary condition $\left. V_{\mathrm{eff}}^{\prime }\left( r\right) \right|_{r=r_{\mathrm{ph}}} = 0 $ yields the equation  
\begin{equation}  
1 - \frac{3M}{r_{\mathrm{ph}}} + \frac{2\eta Q^{2}}{r_{\mathrm{ph}}^{2}} + \frac{16\sqrt{\Theta}M}{\sqrt{\pi}r_{\mathrm{ph}}^{2}} - \frac{10\eta\sqrt{\Theta}Q^{2}}{\sqrt{\pi}r_{\mathrm{ph}}^{3}} = 0.  
\end{equation}  
This equation lacks an analytical solution; hence, the roots are computed numerically. For $ Q = \Theta = 0 $, the photon sphere radius simplifies to $ r_{\mathrm{ph}} = 3M $, which corresponds to the Schwarzschild black hole. In the case of a phantom RN black hole $( \Theta = 0 )$, the photon sphere radius becomes 
\begin{equation}
r_{\mathrm{ph}} = \frac{3M}{2} + \frac{1}{2}\sqrt{9M^{2} - 8\eta Q^{2}}.    
\end{equation}
Table \ref{tbb} displays the numerical results for the photon sphere radius $ r_{\mathrm{ph}} $ and the corresponding impact parameter $ b_{\mathrm{cr}} $ for various values of $ \Theta $ and $ Q $, with $ \eta = \pm 1 $.  
\begin{table}[tbh]  
\centering  
\begin{tabular}{|l|lll|l|lll|}  
\hline\hline  
\rowcolor{lightgray} \multicolumn{4}{l|}{$\eta = 1$ and $\ell = 4$} & \multicolumn{4}{l|}{$\eta = -1$ and $\ell = 4$} \\ \hline  
$\Theta$ & $r_{\mathrm{ph}}$ & $b_{\mathrm{cr}}$ & & $\Theta$ & $r_{\mathrm{ph}}$ & $b_{\mathrm{cr}}$ & \\ \hline\hline  
0.01 & 0.268671 & 0.325968 & & 0.01 & 3.61267 & 1.82550 & \\  
0.03 & 0.430326 & 0.538063 & & 0.03 & 3.39198 & 1.81210 & \\  
0.05 & 0.518517 & 0.651504 & & 0.05 & 3.21720 & 1.80119 & \\  
0.07 & 0.577313 & 0.722641 & & 0.07 & 3.05426 & 1.79182 & \\  
0.09 & 0.620195 & 0.770603 & & 0.09 & 2.89001 & 1.78648 & \\  
0.10 & 0.637706 & 0.788927 & & 0.10 & 2.80385 & 1.62825 & \\  
0.11 & 0.653253 & 0.804472 & & 0.11 & 2.71254 & 1.61178 & \\  
0.12 & 0.667175 & 0.817756 & & 0.12 & 2.61307 & 1.59285 & \\  
0.13 & 0.679737 & 0.829181 & & 0.13 & 2.50009 & 1.57006 & \\  
\hline\hline  
\end{tabular}  
\caption{Photon sphere radius $r_{\mathrm{ph}}$ and impact parameter $b_{\mathrm{cr}}$ for various values of $\Theta$ with $\eta = \pm 1$.}  
\label{tbb}  
\end{table}

\newpage Table \ref{tbb} reveals that, in the ordinary case, the photon sphere radius and the impact parameter increase as the noncommutative parameter $\Theta$ grows. Conversely, in the phantom scenario, they decrease as $\Theta$ increases. This indicates that noncommutativity modifies the spacetime curvature, causing the photon sphere to shift outward in the ordinary scenario and inward in the phantom scenario. These trends underscore the contrasting effects of noncommutative geometry on black holes surrounded by ordinary and phantom fields.

For a black hole spacetime with a (pseudo-)cosmological horizon, such as the Kottler black hole, the size of the black hole's shadow can explicitly vary depending on the radial coordinate of the observer and whether the observer is static or moving with the flow. For a static observer at a distance $r_{\mathrm{O}}$, the angular size of the black hole's shadow, $\alpha_{\mathrm{sh}}$, is determined by the following relation:
\begin{equation}
\sin ^{2}\alpha _{\mathrm{sh}}=\frac{r_{\mathrm{ph}}^{2}}{f\left( r_{\mathrm{%
ph}}\right) }\frac{f\left( r_{\mathrm{O}}\right) }{r_{\mathrm{O}}^{2}} = b_{\mathrm{cr}}^2 \frac{f(r_{\mathrm{O}})}{r_{\mathrm{O}}^2}.
\end{equation}
It is worth noting that this expression is originally derived for asymptotically flat spacetimes, where light rays either escape to infinity or fall into the black hole. In asymptotically AdS spacetimes, the conformal boundary reflects all signals, modifying the global causal structure. However, for an observer at finite $r_{\mathrm{O}}$, the shadow remains primarily determined by the photon sphere, as the critical impact parameter $b_{\mathrm{cr}}$ governs photon capture. In this regime, the AdS boundary has a negligible influence on the shadow's angular size, making the above relation a valid approximation. Significant deviations would arise only for an observer at very large $r_{\mathrm{O}}$, where multiple reflections contribute to the observed image.

\begin{figure}[H]
\begin{minipage}[t]{0.5\textwidth}
        \centering
        \includegraphics[width=\textwidth]{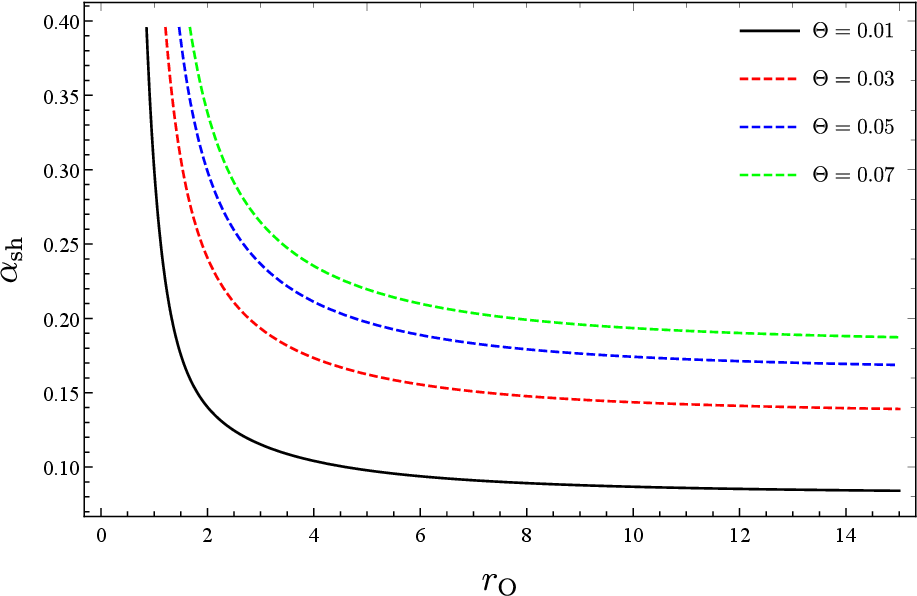}
        \subcaption{$\eta = 1$}
        \label{fig:alphaa}
\end{minipage}
\begin{minipage}[t]{0.5\textwidth}
        \centering
        \includegraphics[width=\textwidth]{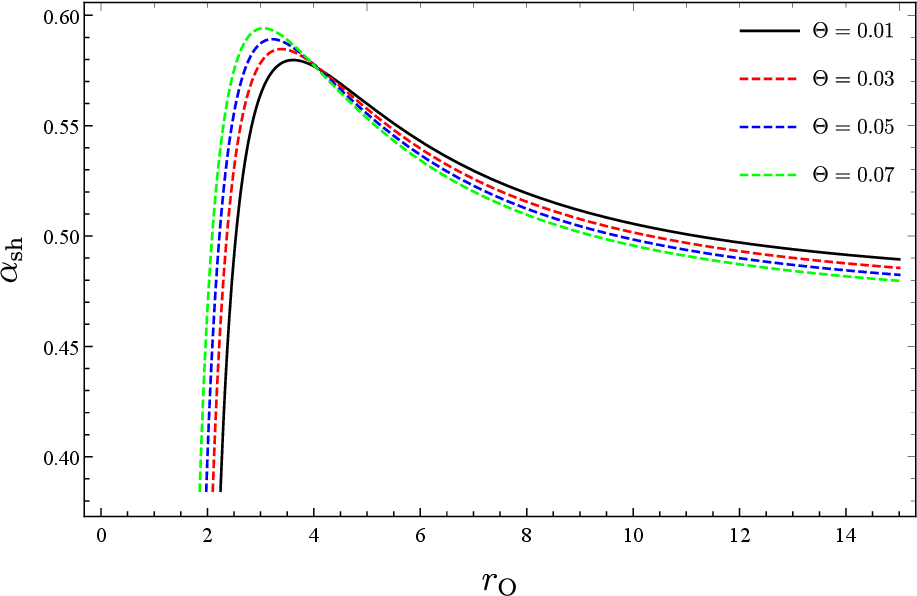}
        \subcaption{$\eta = -1$}
        \label{fig:alphab}
\end{minipage}
\caption{Variation of the angular radius of the black hole's shadow, $\alpha_{\mathrm{sh}}$, with respect to the distance $r_{\mathrm{O}}$ for different values of the $\Theta$ parameter. The used parameters are $M = 1$, $Q = 1.3$, and $\ell = 4$.}
\label{fig:alpha}
\end{figure}

When $r_{\mathrm{O}} = r_{\mathrm{ph}}$, meaning the observer is located at the photon sphere, the angular size of the shadow becomes $\alpha_{\mathrm{sh}} = \pi / 2$, indicating that the shadow occupies exactly half of the observer's sky. The behavior of the static angular radius of the shadow, $\alpha_{\mathrm{sh}}$, as a function of $r_{\mathrm{O}}$ is illustrated in Figure \ref{fig:alpha}.

\noindent Panel (a), for $\eta = 1$, shows that the angular radius decreases as $r_{\mathrm{O}}$ increases, with higher values of $\Theta$ resulting in a steeper decline. In contrast, panel (b), for $\eta = -1$, reveals a more complex behavior: the angular radius initially increases, reaches a maximum, and then decreases for larger values of $r_{\mathrm{O}}$. These observations highlight the significant influence of both $\Theta$ and $\eta$ on the black hole's shadow, emphasizing their importance in determining the observable features of the black hole.

On the other hand, the shadow radius of a black hole for a static observer at a given position $ r_{\mathrm{O}} $ is given by:
\begin{eqnarray}
R_{\mathrm{sh}} &=&r_{\mathrm{ph}}\sqrt{\frac{1-\frac{2M}{r_{\mathrm{O}}}+\eta \frac{Q^{2}}{r_{\mathrm{O}}^{2}}+\frac{r_{\mathrm{O}}^{2}}{\ell ^{2}}}{1-\frac{2M}{r_{\mathrm{ph}}}+\eta \frac{Q^{2}}{r_{\mathrm{ph}}^{2}}+\frac{r_{
\mathrm{ph}}^{2}}{\ell ^{2}}}}+2\sqrt{\frac{\Theta }{\pi }}\frac{r_{\mathrm{%
ph}}\left( \frac{2M}{r_{\mathrm{O}}^{2}}-\eta \frac{Q^{2}}{r_{\mathrm{O}}^{3}%
}\right) }{\sqrt{\left( 1-\frac{2M}{r_{\mathrm{ph}}}+\eta \frac{Q^{2}}{r_{%
\mathrm{ph}}^{2}}+\frac{r_{\mathrm{ph}}^{2}}{\ell ^{2}}\right) \left( 1-%
\frac{2M}{r_{\mathrm{O}}}+\eta \frac{Q^{2}}{r_{\mathrm{O}}^{2}}+\frac{r_{%
\mathrm{O}}^{2}}{\ell ^{2}}\right) }}  \notag \\
&&-2\sqrt{\frac{\Theta }{\pi }}\frac{\frac{2M}{r_{\mathrm{ph}}}-\eta \frac{%
Q^{2}}{r_{\mathrm{ph}}^{2}}}{\left( 1-\frac{2M}{r_{\mathrm{ph}}}+\eta \frac{%
Q^{2}}{r_{\mathrm{ph}}^{2}}+\frac{r_{\mathrm{ph}}^{2}}{\ell ^{2}}\right) }%
\sqrt{\frac{1-\frac{2M}{r_{\mathrm{O}}}+\eta \frac{Q^{2}}{r_{\mathrm{O}}^{2}}%
+\frac{r_{\mathrm{O}}^{2}}{\ell ^{2}}}{1-\frac{2M}{r_{\mathrm{ph}}}+\eta 
\frac{Q^{2}}{r_{\mathrm{ph}}^{2}}+\frac{r_{\mathrm{ph}}^{2}}{\ell ^{2}}}} 
\notag \\
&&-2\frac{\Theta r_{\mathrm{ph}}}{\pi r_{\mathrm{O}}^{2}}\frac{\left( \frac{%
2M}{r_{\mathrm{O}}}-\eta \frac{Q^{2}}{r_{\mathrm{O}}^{2}}\right) ^{2}}{%
\left( 1-\frac{2M}{r_{\mathrm{O}}}+\eta \frac{Q^{2}}{r_{\mathrm{O}}^{2}}+%
\frac{r_{\mathrm{O}}^{2}}{\ell ^{2}}\right) ^{\frac{3}{2}}\left( 1-\frac{2M}{%
r_{\mathrm{ph}}}+\eta \frac{Q^{2}}{r_{\mathrm{ph}}^{2}}+\frac{r_{\mathrm{ph}%
}^{2}}{\ell ^{2}}\right) }  \notag \\
&&+\frac{6\Theta }{\pi r_{\mathrm{ph}}}\frac{\left( \frac{2M}{r_{\mathrm{ph}}%
}-\eta \frac{Q^{2}}{r_{\mathrm{ph}}^{2}}\right) ^{2}}{\left( 1-\frac{2M}{r_{%
\mathrm{ph}}}+\eta \frac{Q^{2}}{r_{\mathrm{ph}}^{2}}+\frac{r_{\mathrm{ph}}^{2}}{\ell ^{2}}\right) ^{2}}\sqrt{\frac{1-\frac{2M}{r_{\mathrm{O}}}+\eta\frac{Q^{2}}{r_{\mathrm{O}}^{2}}+\frac{r_{\mathrm{O}}^{2}}{\ell ^{2}}}{1-\frac{2M}{r_{\mathrm{ph}}}+\eta \frac{Q^{2}}{r_{\mathrm{ph}}^{2}}+\frac{r_{\mathrm{ph}}^{2}}{\ell ^{2}}}}.  \label{shad}
\end{eqnarray}%
It is evident that for a static observer at a large distance, i.e., as $r_{\mathrm{O}}\rightarrow \infty $, the shadow radius $ R_{\mathrm{sh}} $ does not approach the critical value $ b_{\mathrm{cr}} $.

Let us now establish the constraints on the non-commutative factor by analyzing the observational data from the EHT. Using the EHT horizon scale for SgrA$^{\ast}$, the mass-to-distance ratio priors from Keck and VLTI are averaged. By considering the bounds mentioned in \cite{Sunny}, the shadow radius $R_{\mathrm{sh}}$ is constrained within the following ranges:
\begin{itemize}
\item For $ 1\sigma $ constraints:
\begin{equation}
4.55 \leq \frac{R_{\mathrm{sh}}}{M} \leq 5.22,    
\end{equation}
\item For $ 2\sigma $ constraints:
\begin{equation}
4.21 \leq \frac{R_{\mathrm{sh}}}{M} \leq 5.56.
\end{equation}
\end{itemize}
To simplify the calculation, we consider the Schwarzschild black hole in non-commutative space. In this case, the equation for the photonic radius is given by:
\begin{equation}
1 - \frac{3M}{r_{\mathrm{ph}}} + \frac{16\sqrt{\Theta}M}{\sqrt{\pi}r_{\mathrm{ph}}^2} = 0. \label{sd}
\end{equation}
Solving Eq. (\ref{sd}) yields:
\begin{equation}
r_{\mathrm{ph}} \simeq 3M - \frac{16}{3M} \sqrt{\frac{\Theta}{\pi}}. \label{sd1}
\end{equation}
By substituting Eq. (\ref{sd1}) into the expression for the shadow radius, we obtain:
\begin{equation}
\frac{R_{\mathrm{sh}}}{M} = 3\sqrt{3} - 4\sqrt{\frac{3\hat{\Theta}}{\pi}},
\end{equation}
where $\hat{\Theta} = \frac{\Theta}{M}$.

\begin{figure}[H]
\begin{minipage}[t]{0.5\textwidth}
        \centering
        \includegraphics[width=\textwidth]{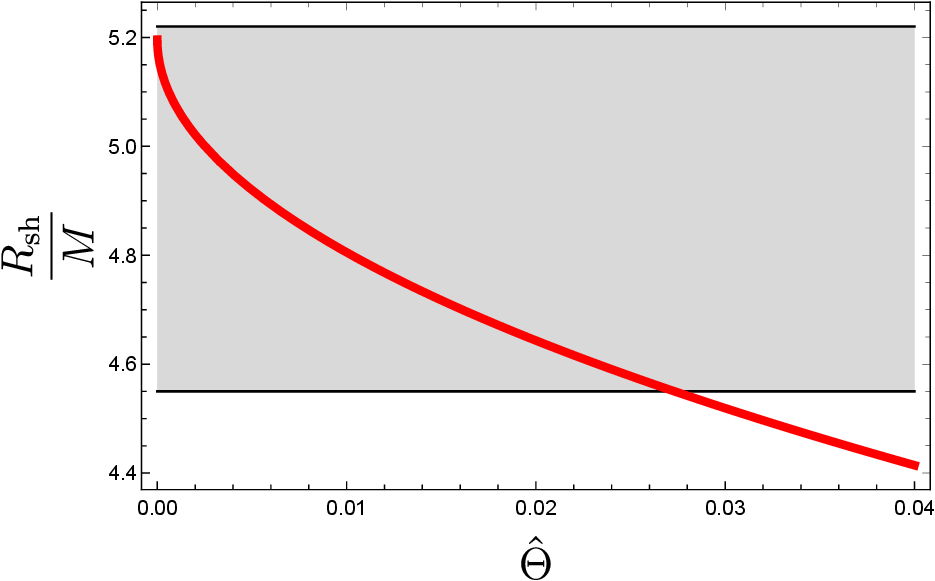}
                \subcaption{$1\sigma$}
        \label{fig:sigma1}
\end{minipage}
\begin{minipage}[t]{0.5\textwidth}
        \centering
        \includegraphics[width=\textwidth]{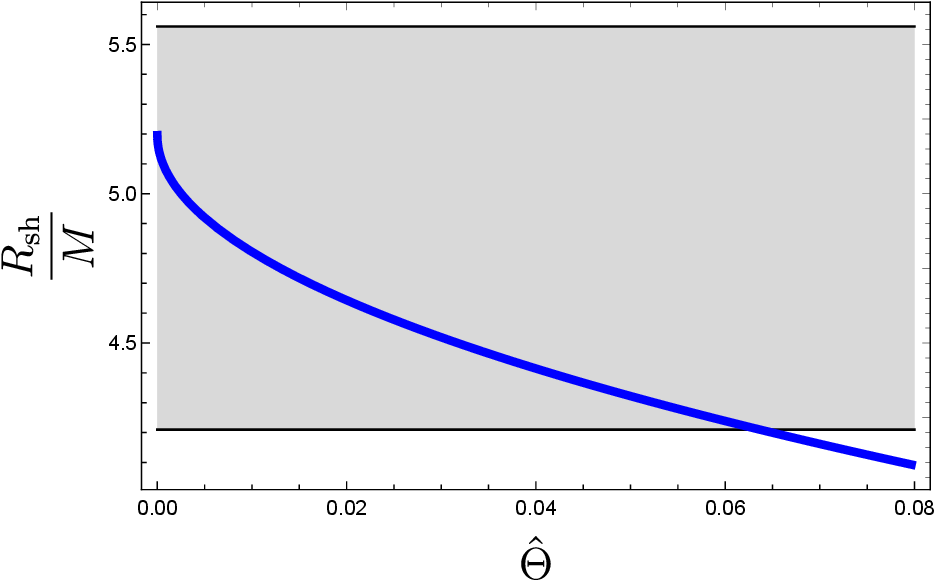}
                \subcaption{$2\sigma$}
        \label{fig:sigma2}
\end{minipage}
\caption{Shadow radius $\frac{R_{\mathrm{sh}}}{M}$ of the Schwarzschild black hole in noncommutative space, expressed in units of mass, as a function of $\hat{\Theta}$.}
\label{fig:sigma}
\end{figure}

Figure \ref{fig:sigma} shows the shadow radius, expressed in units of mass, as a function of the noncommutativity parameters. 

\noindent The plot intersects the experimental constraint lines at two points, providing an upper limit on the NC parameter. Based on the first set of experimental data, we find $\hat{\Theta} \leq 0.027 $, and from the second set, we obtain $ \hat{\Theta} \leq 0.063$.

\section{Quasinormal modes} \label{sec6}



In this section, we compute the QNMs of a test scalar field in the background of asymptotically AdS spherically symmetric noncommutative RN-black holes. The scalar field $\Phi$ in a curved spacetime satisfies the general covariant Klein-Gordon equation:
\begin{equation}\label{coveqs}
\frac{1}{\sqrt{-g}} \partial_\mu \left(\sqrt{-g} g^{\mu \nu} \partial_\nu \Phi \right) = 0,
\end{equation}
After separation of variables, Eq.~(\ref{coveqs}) can be reduced to a wave-like form with effective potentials \cite{Kokkotas:1999bd,Berti:2009kk,Konoplya:2011qq}:
\begin{equation}\label{wave-equation}
\frac{d^2 \Psi}{dr_*^2} + \left(\omega^2 - V_s(r)\right)\Psi = 0,
\end{equation}
where the ``tortoise coordinate'' $r_*$ is defined as
\begin{equation}\label{tortoise}
\frac{dr_*}{dr} \equiv \frac{1}{f(r)},
\end{equation}
and the effective potential takes the form
\begin{equation}\label{potentialScalar}
V(r) = f(r) \frac{l(l+1)}{r^2} + \frac{1}{r} \frac{d^2 r}{dr_*^2},
\end{equation}
where $l = 0, 1, 2, \ldots$ is the multipole number.

\begin{figure} [htbp]
\resizebox{\linewidth}{!}{\includegraphics{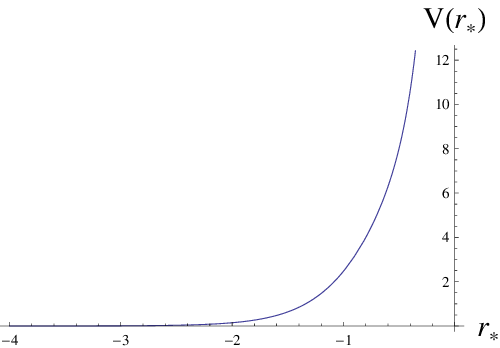}\includegraphics{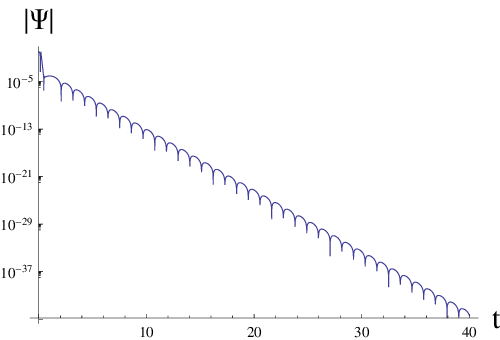}\includegraphics{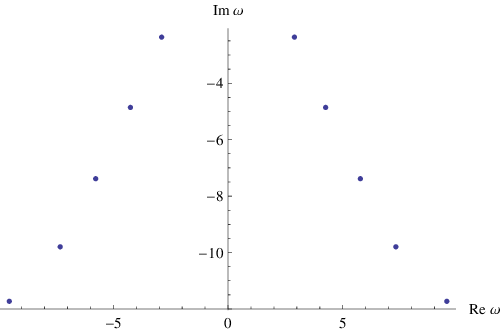}}
\caption{Effective potential as a function of the tortoise coordinate (left), time-domain profile (middle) and first five quasinormal modes for $l=1$, $M=1$, $Q=0$, $\eta =0$, $\Theta =0.01$, $\ell=1$. }\label{fig:QNM1}
\end{figure}

\begin{figure} [htbp]
\resizebox{\linewidth}{!}{\includegraphics{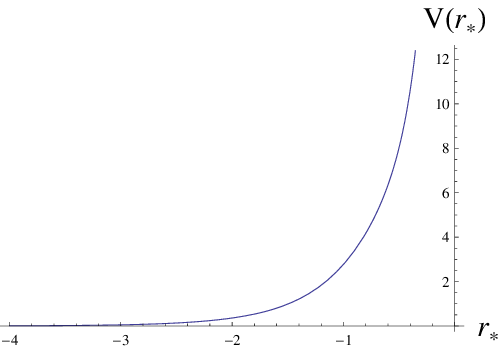}\includegraphics{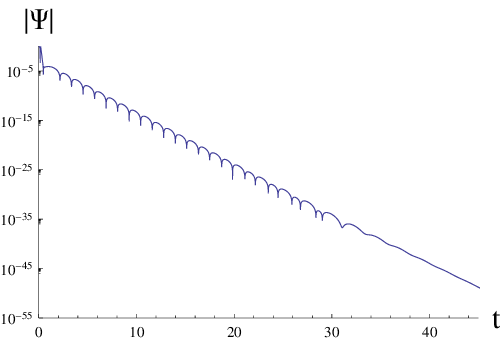}\includegraphics{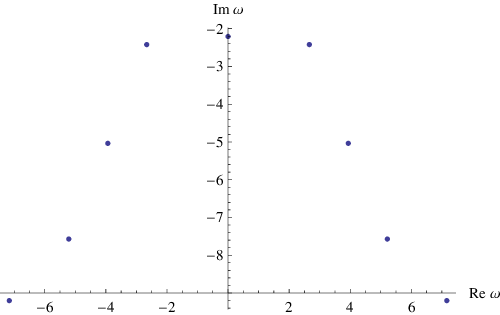}}
\caption{Effective potential as a function of the tortoise coordinate (left), time-domain profile (middle) and first five quasinormal modes for $l=1$, $M=1$, $Q=0.1$, $\eta =1$, $\Theta =0.02$, $\ell=1$. }\label{fig:QNM2}
\end{figure}

For asymptotically AdS spacetimes, the effective potential diverges at the AdS boundary (see Figs.~\ref{fig:QNM1} and \ref{fig:QNM2}), which implies that the AdS spacetime acts as an effective confining box. Therefore, Dirichlet boundary conditions are imposed at the AdS boundary, and purely ingoing waves are imposed at the event horizon \cite{Horowitz:1999jd}.
\begin{equation}
\Psi   \sim e^{i \omega (t+r^{*})}, \quad r \rightarrow r_{H} \
\quad (r^{*} \rightarrow -\infty)
\end{equation}
\begin{equation}
\Psi \rightarrow 0, \quad r \rightarrow \infty \quad (r^{*} \quad is \quad finite)
\end{equation}
We calculate the QNMs of the noncommutative RN-AdS black hole for the ordinary and phantom cases and tabulate it in Table  \ref{table4}.

\begin{table} [tbh]
    \centering
\begin{tabular}{|l|l|}
  \hline
   \hline
 \rowcolor{lightgray} Parameters & $\omega$  \\ \hline
  \hline
  $Q^{2} \eta = 0$, $\Theta = 0$,  \qquad $l=0$ & $2.798244 - 2.670872 i$, \quad $4.758579 - 5.036394 i$ \\
  $Q^{2} \eta = 0$, $\Theta = 0.01$, \,\, $l=0$ & $2.235784 - 2.701012 i$, \quad $3.841361 - 5.275000 i$ \\
  $Q^{2} \eta = 0$, $\Theta = 0.02$, \,\, $l=0$ & $2.176148 - 2.871816 i$, \quad $4.313182 - 5.309921 i$ \\
  $Q^{2} \eta = 0$, $\Theta = 0$, \quad  \,\,\,\, $l=1$ &   $3.330557 - 2.489059 i$, \quad $5.172406 - 4.886582 i$ \\
  $Q^{2} \eta = 0$, $\Theta = 0.01$, \,\,\,  $l=1$ & $2.892727 - 2.367050 i$, \quad $4.250723 - 4.858340 i$ \\
  $Q^{2} \eta = 0$, $\Theta = 0.02$, \,\,\, $l=1$ & $2.670853 - 2.423699 i$, \quad$3.954439 - 5.034608 i$ \\
    \hline
  $Q^{2} \eta = 0.01$, $\Theta = 0.02$, $l=1$ & $- 2.207833 i$, \quad $2.662244 - 2.421744 i$, \quad $3.935758 - 5.034321 i$ \\
  $Q^{2} \eta = 0.04$, $\Theta = 0.02$, $l=1$ & $-1.937026 i$, \quad $2.635799 -2.417316 i$, \quad $3.887119 - 5.032424 i$ \\
  $Q^{2} \eta = 0.09$, $\Theta = 0.02$, $l=1$ & $ - 1.434130 i$, \quad $2.590320 - 2.411070 i$, \quad $3.789750 - 5.024180 i$ \\
  $Q^{2} \eta = 0.16$, $\Theta = 0.02$, $l=1$ & $- 0.238302 i$, \, $-0.396832 i$, \quad $2.52354 - 2.40588 i$, \quad $-0.555748 i$ \\
  \hline
   \hline
\end{tabular}
\caption{The fundamental mode and the first overtones for various values of parameters. Here we have $M=1$ and $\ell=1$.  At larger $Q^2 \eta$ the new non-oscillatory modes dominate in the late time signal.}\label{table4}
\end{table}

The discretization scheme for time-domain integration, which is more efficient for asymptotically AdS spacetimes, is given by the following form \cite{Wang:2004bv}:
\begin{equation}
\begin{aligned}
    \left[ 1 + \frac{\Delta^2}{16} V(S) \right] \psi(N) &= \psi(E) + \psi(W) - \psi(S) \\
    &\quad - \frac{\Delta^2}{16} \big[ V(S)\psi(S) + V(E)\psi(E) \big] + V(W)\psi(W).
\end{aligned}
\end{equation}
Here, the points $N$, $S$, $W$, and $E^2$ are defined as follows: $N = (u + \Delta, v + \Delta)$, $W = (u + \Delta, v)$, $E = (u, v + \Delta)$, and $S = (u, v)$. The truncation error associated with this scheme is of the order $\mathcal{O}(\Delta^4)$. This discretization scheme has also been applied in the analysis of the evolution of perturbations for other black hole models, including those involving phantom fields \cite{Bronnikov2012}.

\newpage
Quasinormal spectra of asymptotically AdS black holes differ qualitatively from those of asymptotically flat black holes. The main features of the asymptotically AdS spectrum are as follows:
\begin{itemize}
    \item The spectrum quickly becomes equidistant in both the real and imaginary parts.
    \item As the mass of the black hole approaches zero (or equivalently, as the ratio of the event horizon radius to the AdS radius approaches zero), the quasinormal modes of the black hole converge to those of empty AdS space \cite{Konoplya:2002zu,Cardoso:2003cj}, unless an instability occurs for sufficiently small black holes \cite{Konoplya:2017zwo,Konoplya:2017ymp}.
    \item For large black holes, the quasinormal mode frequencies scale with the black hole radius \cite{Horowitz:1999jd}.
    \item Quasinormal modes govern the decay of perturbations not only at intermediate (ringdown) times but also at asymptotic times \cite{Konoplya:2007jv,Kooplya2008}. This behavior is also observed for the asymptotically de Sitter case \cite{Dyatlov:2010hq,Dyatlov:2011jd}, and it has a practical consequence: not only the fundamental mode but also a few overtones can be extracted from time-domain profiles. This is because the ringdown is not contaminated by asymptotic power-law tails at late times. We will use this property to find overtones.
\end{itemize}

From Fig.~\ref{fig:QNM2} and Table~\ref{table4}, it is evident that when all parameters $\eta$, $\Theta$, and $Q$ are nonzero, a qualitatively new mode appears, characterized by a zero real part. This mode is, therefore, non-oscillatory and likely represents a new branch of modes that is absent in the spectrum of a scalar field in the background of Schwarzschild-anti-de Sitter black holes \cite{Horowitz:1999jd,Cardoso:2003cj,Konoplya:2002zu}.

We interpret the nonzero values of $\Theta$ and $Q \eta$ as representing a static deformation of the Schwarzschild-anti-de Sitter metric near the event horizon, while maintaining the same asymptotic behavior in the far zone. Consequently, the appearance of this new branch of modes can be viewed as a result of static deformations in the near-horizon metric, consistent with the phenomenon of an outburst of overtones observed for asymptotically AdS black holes and branes in \cite{Konoplya:2023kem}. 

A similar phenomenon, demonstrating the high sensitivity of spectra to near-horizon deformations, has recently been observed for various asymptotically flat black holes \cite{Zinhailo:2024kbq,Bolokhov:2023bwm,Luo:2024dxl}, allowing the first few overtones of quasinormal modes to be interpreted as "the sound of the event horizon" \cite{Konoplya:2022pbc,Konoplya:2022hll}. Thus, we observe that even relatively small nonzero values of charge, the phantom parameter, and the parameters of non-commutativity lead to significant changes in the quasinormal spectrum, resulting in the emergence of qualitatively new, non-oscillatory modes.

It is worth noting that, as the first study of the quasinormal spectrum of phantom RN-AdS non-commutative black holes, we have analyzed the massless scalar field. However, this analysis could be extended in a similar manner to other fields, such as electromagnetic, Dirac, and gravitational fields.

\bigskip

\section{Conclusion} \label{sec7}

In this work, we investigated the influence of noncommutativity on phantom Reissner-Nordström-Anti-de Sitter (RN-AdS) black holes, focusing on their geometric, thermodynamic, and observational properties. By incorporating Lorentzian mass and charge distributions, we derived a modified metric function that exhibits significant deviations from the classical RN-AdS case. These modifications alter the horizon structure, leading to the suppression of singularities and shifting the event horizon positions depending on the noncommutative parameter $ \Theta $.

A thermodynamic analysis was conducted, where we examined the Hawking temperature, entropy, heat capacity, and phase transitions of the black hole. Our findings indicate that noncommutativity influences black hole stability, modifies the Gibbs free energy profile, and alters the phase transition structure, particularly in the phantom case. Notably, noncommutative effects delay the onset of thermodynamic stability and introduce corrections to the entropy beyond the standard Bekenstein-Hawking relation.

We further explored the black hole’s role as a heat engine, demonstrating that noncommutativity enhances the efficiency of the thermodynamic cycle. However, for phantom RN-AdS black holes, while efficiency is improved, it remains lower than in the ordinary case due to the exotic nature of phantom fields. Additionally, we analyzed the motion of test particles and photons, deriving the effective potential, innermost stable circular orbits, and the black hole shadow profile. We found that increasing $ \Theta $ alters the potential barrier and shifts the ISCO radius, with distinct behaviors for standard and phantom fields.

Finally, we computed quasinormal modes of a test scalar field to assess the dynamical stability of the black hole. Our results reveal that noncommutativity modifies the damping rates and introduces non-oscillatory modes that are absent in the classical case. The transition from oscillatory to purely decaying modes in the near-extremal regime suggests a novel late-time response, providing potential observational signatures.

Overall, our results highlight the intricate interplay between noncommutativity and phantom fields, revealing their combined effects on black hole structure, thermodynamics, stability, and observables. These findings contribute to the broader understanding of quantum corrections in black hole physics and may provide insights into testing alternative gravity models through astrophysical observations.

\section*{Acknowledgments}
The authors sincerely thank R. A. Konoplya for valuable scientific discussions, particularly regarding the quasinormal mode analysis, which significantly contributed to this work. B. C. L. is grateful to Excellence Project PřF UHK 2205/2025-2026 for the financial support.

\section*{Data Availability Statement}
The manuscript has no associated data.


\begin{thebibliography}{999}

\bibitem{Reissner} H. Reissner, \href{https://doi.org/10.1002/andp.19163550905}{Ann. Phys. (Berlin),  \textbf{50}, 106 (1916).}

\bibitem{Weyl} H. Weyl, \href{https://doi.org/10.1002/andp.19173591804}{Ann. Phys. (Berlin), \textbf{54}, 117 (1917).} 

\bibitem{Nordstrom} G. Nordström, \href{https://dwc.knaw.nl/DL/publications/PU00012316.pdf}{Proc. Kon. Ned. Akad. Wet. \textbf{20}, 1238 (1918).}

%
\bibitem{Jeffery} G. B. Jeffery, \href{https://royalsocietypublishing.org/doi/abs/10.1098/rspa.1921.0028}{Proc. R. Soc. Lond. A. \textbf{99}, 123 (1921).}


\bibitem{Louko1996} J. Louko, S. N. Winters-Hilt, \href{https://doi.org/10.1103/PhysRevD.54.2647}{Phys. Rev. D \textbf{54}, 2647 (1996).}

\bibitem{Witten1998} E. Witten, \href{https://doi.org/10.4310/ATMP.1998.v2.n2.a2}{Adv. Theor. Math. Phys. \textbf{2}, 253 (1998).} 

\bibitem{Chamblin1999} A. Chamblin, R. Emparan, C. V. Johnson, R. C. Myers, \href{https://doi.org/10.1103/PhysRevD.60.104026}{Phys. Rev. D \textbf{60}, 104026 (1999).}

\bibitem{Maldacena1999} J. Maldacena, \href{https://doi.org/10.1023/A:1026654312961}{Int. J. Theor. Phys. \textbf{38}, 1113 (1999).}


\bibitem{Wang2000} B. Wang, C. Y. Lin, E. Abdalla, \href{https://doi.org/10.1016/S0370-2693(00)00409-3}{Phys. Lett. B \textbf{481}, 79 (2000).}





\bibitem{Bekenstein1973} J. Bekenstein, \href{https://doi.org/10.1103/PhysRevD.7.2333}{Phys. Rev. D \textbf{7}, 2333 (1973).}


\bibitem{Hawking1975} S. W. Hawking, \href{https://doi.org/10.1007/BF02345020}{Commun. Math. Phys. \textbf{43}, 199 (1975).}


\bibitem{Bardeen1973} J. M. Bardeen, B. Carter, S. W. Hawking, \href{ https://doi.org/10.1007/BF01645742}{Commun. Math. Phys. \textbf{31},  161 (1973).}


\bibitem{Hawking1983} S. W. Hawking, D. N. Page, \href{ https://doi.org/10.1007/BF01208266}{Commun. Math. Phys. \textbf{87}, 577 (1983).}


\bibitem{Brown1994} J. D. Brown, J. Creighton, R. B. Mann, \href{https://doi.org/10.1103/PhysRevD.50.6394}{Phys. Rev. D \textbf{50}, 6394 (1994).}


\bibitem{Cai1996} R. G. Cai, Y. Z. Zhang, \href{https://doi.org/10.1142/S0217732396002010}{Mod. Phys. Lett. A \textbf{11}, 2027 (1996).}


\bibitem{Peca1999} C. S. Peca, J. P. S. Lemos, \href{https://doi.org/10.1103/PhysRevD.59.124007}{Phys. Rev. D \textbf{59}, 124007 (1999).}

\bibitem{Dolan2011} B. P. Dolan,\href{https://doi.org/10.1088/0264-9381/28/23/235017}{Class. Quant. Grav. \textbf{28}, 235017 (2011).}


\bibitem{Kubiznak2012} D. Kubizňák, R. B. Mann, \href{https://doi.org/10.1007/JHEP07(2012)033}{J. High Energy Phys. \textbf{2012}, 033 (2012).}


\bibitem{Kubiznak2017} D. Kubizňák, R. B. Mann, M. Teo, \href{https://doi.org/10.1088/1361-6382/aa5c69}{Class. Quantum Grav. \textbf{34}, 063001 (2017).}

\bibitem{Hamil20241} B. Hamil,   B. C. L\"{u}tf\"{u}o\u{g}lu, L. Dahbi,
\href{https://dx.doi.org/10.1142/S021773232450161X}{Mod. Phys. Lett. A \textbf{39}, 2450161 (2024).} 


\bibitem{Kooplya2008} R. A. Konoplya, A. Zhidenko, \href{https://doi.org/10.1103/PhysRevD.78.104017}{Phys. Rev. D \textbf{78}, 104017 (2008).}


\bibitem{Niu2012} C. Niu, Y. Tian, X. N. Wu, \href{https://doi.org/10.1103/PhysRevD.85.024017}{Phys. Rev. D \textbf{85}, 024017 (2012).}

\bibitem{Liu2014} Y. Liu, D. C. Zou,  B. Wang, \href{https://doi.org/10.1007/JHEP09(2014)179}{J. High Energ. Phys. \textbf{2014}, 179 (2014).}  


\bibitem{Chen2019} D. Y. Chen, W. Yang, X. X. Zeng, \href{https://doi.org/10.1016/j.nuclphysb.2019.11472}{Nucl. Phys. B \textbf{946}, 114722 (2019).}



\bibitem{Li2020} R. Li, K. Zhang, J. Wang, \href{https://doi.org/10.1007/JHEP10(2020)090}{J. High Energ. Phys. \textbf{2020}(10), 90 (2020).}


\bibitem{Zhang2021} J. Zhang, Y. Y. Lin, H. C. Liang, K. J. Chi, C. M. Chen, S. P. Kim, J. R. Sun, \href{https://doi.org/10.1088/1674-1137/abf4f6}{Chinese Phys. C \textbf{45}, 065105 (2021).}


\bibitem{Ghaffarnejad2025} H. Ghaffarnejad, E. Ghasemi, \href{https://doi.org/10.1007/s10714-025-03374-5}{Gen. Relativ. Gravit. \textbf{57}, 37 (2025).}  


\bibitem{M871} K. Akiyama et al [Event Horizon Telescope Collaboration et al.], \href{https://doi.org/10.3847/2041-8213/ab0ec7}{
Astrophys. J. Lett. \textbf{875}, L1 (2019).}

\bibitem{SagA1} K. Akiyama et al [Event Horizon Telescope Collaboration et al.], \href{https://doi.org/10.3847/2041-8213/ac6674}{
Astrophys. J. Lett. \textbf{930}, L12 (2022).}



\bibitem{Atamurotov2013} F. Atamurotov, A. Abdujabbarov, B. Ahmedov, \href{https://doi.org/10.1103/PhysRevD.88.064004}{Phys. Rev. D \textbf{88}, 064004 (2013).}


\bibitem{Roman2019} R. A. Konoplya, \href{https://doi.org/10.1016/j.physletb.2019.05.043}{Phys. Lett. B \textbf{795}, 1 (2019).}

\bibitem{Guo2020} M. Guo, P. C. Li, \href{https://doi.org/10.1140/epjc/s10052-020-8164-7}{Eur. Phys. J. C \textbf{80}, 588 (2020).}   

\bibitem{Anacleto2021} M. A. Anacleto, J. A. V. Campos,
F. A. Brito, E. Passos, \href{https://doi.org/10.1016/j.aop.2021.168662}{Ann. Phys. \textbf{434}, 168662 (2021).} 

\bibitem{Pantig2022} R. C. Pantig, P. K. Yu, E. T. Rodulfo, A. \"Ovg\"un, \href{https://doi.org/10.1016/j.aop.2021.168722}{Ann. Phys. \textbf{436}, 168722 (2022).} 

\bibitem{Okyay2022} M. Okyay, A. \"Ovg\"un, \href{https://doi.org/10.1088/1475-7516/2022/01/009}{J. Cosmol. Astropart. Phys. \textbf{01}, 009 (2022).}

\bibitem{Hamil20231} B. Hamil, B. C.  L\"{u}tf\"{u}o\u{g}lu, \href{https://doi.org/10.1016/j.dark.2023.101293}{
Phys. Dark Universe \textbf{42}, 101293 (2023).}

\bibitem{Hamil20232} B. Hamil, B. C. L\"{u}tf\"{u}o\u{g}lu, \href{https://doi.org/10.1016/j.nuclphysb.2023.116191}{Nucl. Phys. B \textbf{990}, 116191  (2023).}



\bibitem{Hamil20243} B. Hamil, B. C. L\"{u}tf\"{u}o\u{g}lu, \href{https://doi.org/10.1088/1674-1137/ad2a4d}{Chinese Phys. C \textbf{48}(05), 055102 (2024).}


\bibitem{Yunusov2024} O. Yunusov, J. Rayimbaev, F. Sarikulov, M. Zahid, A. Abdujabbarov, Z. Stuchlik, \href{https://doi.org/10.1140/epjc/s10052-024-13500-3}{Eur. Phys. J. C \textbf{84}, 1240 (2024).} 



\bibitem{Luo2024} S. Luo, C. H. Li, \href{https://doi.org/10.1103/PhysRevD.110.124042}{Phys. Rev. D \textbf{110}, 124042 (2024).}




\bibitem{Hamil20252} B. Hamil,  B. C. L\"{u}tf\"{u}o\u{g}lu, \href{ https://doi.org/10.1002/prop.202400105}{Fort. der Phys. \textbf{early access} (2024).}



\bibitem{Hamil20251} B. Hamil,  B. C. L\"{u}tf\"{u}o\u{g}lu, \href{https://doi.org/10.1088/1674-1137/ad9894}{Chinese Phys. C \textbf{49}, 025107 (2025).}

\bibitem{Hamil20253} B. Hamil,  B. C. L\"{u}tf\"{u}o\u{g}lu, \href{https://doi.org/10.1016/j.aop.2024.169861}{Ann. Phys.   \textbf{472}, 169861 (2025).}




\bibitem{Belhaj2020} A. Belhaj, L. Chakhchi, H. El Moumni, J. Khalloufi, K. Masmar, \href{https://doi.org/10.1142/S0217751X20501705}{Int. J. Mod. Phys. A \textbf{35}, 2050170 (2020).}


\bibitem{Wang2022} C. Wang, B. Wu, Z. M. Xu, W. L. Yang, \href{https://doi.org/10.1016/j.nuclphysb.2022.115698}{Nucl. Phys. B \textbf{976}, 115698 (2022).}

\bibitem{Mandal2023} S. Mandal, Y. Myrzakulov, G. Yergaliyeva, \href{https://doi.org/10.1142/S0217751X23500471}{Int. J. Mod. Phys. A \textbf{38}, 2350047 (2023).}

\bibitem{Ladino2024} J. M. Ladino, C. E. Romero-Figueroa, H. Quevedo, \href{https://doi.org/10.1016/j.nuclphysb.2024.116734}{Nucl. Phys. B \textbf{1009}, 116734 (2024).}

\bibitem{AraujoFilho2025} A. A. Araújo Filho, \href{https://doi.org/10.1088/1475-7516/2025/01/072}{J. Cosmol. Astropart. Phys. \textbf{01}, 072 (2025).}



\bibitem{Kokkotas:1999bd}
K.~D.~Kokkotas,  B.~G.~Schmidt, \href{https://doi.org/10.12942/lrr-1999-2}{Living Rev. Rel. \textbf{2}, 2 (1999)}

\bibitem{Berti:2009kk}
E.~Berti, V.~Cardoso, A.~O.~Starinets,
\href{https://doi.org/10.1088/0264-9381/26/16/163001}{Class. Quant. Grav. \textbf{26}, 163001 (2009).} 


\bibitem{Konoplya:2011qq}
R.~A.~Konoplya, A.~Zhidenko,
\href{https://doi.org/10.1103/RevModPhys.83.793}{Rev. Mod. Phys. \textbf{83}, 793 (2011).}

\bibitem{Konoplya2002a} R. A. Konoplya, \href{https://doi.org/10.1103/PhysRevD.66.084007}{Phys. Rev. D \textbf{66}, 084007 (2002).}


\bibitem{Press1971} W. H. Press, \href{https://doi.org/10.1086/180849}{Astrophys. J. Lett.  \textbf{170},  L105 (1971).}


\bibitem{Aharony:1999ti}
O.~Aharony, S.~S.~Gubser, J.~M.~Maldacena, H.~Ooguri, Y.~Oz, \href{https://doi.org/10.1016/S0370-1573(99)00083-6}{Phys. Rept. \textbf{323}, 183 (2000).}




\bibitem{Starinets:2002br}
A.~O.~Starinets,
\href{https://doi.org/10.1103/PhysRevD.66.124013}{Phys. Rev. D \textbf{66}, 124013 (2002).}





\bibitem{Kovtun:2005ev}
P.~K.~Kovtun, A.~O.~Starinets,
\href{https://doi.org/10.1103/PhysRevD.72.086009}{ Phys. Rev. D \textbf{72}, 086009 (2005).}

\bibitem{Konoplya:2007jv}
R.~A.~Konoplya, A.~Zhidenko, \href{https://doi.org/10.1016/j.nuclphysb.2007.04.016}{Nucl. Phys. B \textbf{777}, 182 (2007).}

\bibitem{Dyatlov:2010hq}
S.~Dyatlov, \href{https://doi.org/10.1007/s00220-011-1286-x}{Commun. Math. Phys. \textbf{306}, 119 (2011).}

\bibitem{Dyatlov:2011jd}
S.~Dyatlov, \href{https://doi.org/10.1007/s00023-012-0159-y}{Ann. Henri Poincare \textbf{13}, 1101 (2012).}

\bibitem{Konoplya:2017zwo}
R.~A.~Konoplya, A.~Zhidenko,
\href{https://doi.org/10.1007/JHEP09(2017)139}{J. High Energ. Phys. \textbf{09}, 139 (2017).}






\bibitem{BF1} B. P. Abbott et al. [LIGO Scientific and Virgo Collaborations], \href{https://doi.org/10.1103/PhysRevLett.116.061102}{Phys. Rev. Lett. \textbf{116}, 061102 (2016).}

\bibitem{BF2} B. P. Abbott et al. [LIGO Scientific and Virgo Collaborations], \href{https://doi.org/10.1103/PhysRevLett.119.161101}{Phys. Rev. Lett. \textbf{119}, 161101 (2017).}

\bibitem{BF3} E. Barausse, E. Berti, T. Hertog, \href{https://doi.org/10.1007/s10714-020-02691-1}{Gen. Relativ. Gravit. \textbf{52}, 81 (2020).}

  









\bibitem{Dubinsky:2024gwo}
A.~Dubinsky, \href{https://doi.org/10.1142/S0217732324501086}{Mod. Phys. Lett. A \textbf{39}, 2450108 (2024).}


\bibitem{Bolokhov:2023bwm}
S.~V.~Bolokhov,
\href{https://doi.org/10.1103/PhysRevD.110.024010}{Phys. Rev. D \textbf{110}, 024010 (2024).}


\bibitem{Luo:2024dxl}
S.~Luo, \href{https://doi.org/10.1103/PhysRevD.110.084071}{Phys. Rev. D \textbf{110}, 084071 (2024).}

\bibitem{Konoplya:2022pbc}
R.~A.~Konoplya, A.~Zhidenko, \href{https://doi.org/10.1016/j.jheap.2024.10.015}{J.  High Energy Astrophys. \textbf{44}, 419 (2024).}


\bibitem{Zinhailo:2024kbq}
A.~F.~Zinhailo, \href{https://doi.org/10.13140/RG.2.2.26785.01124}{doi:10.13140/RG.2.2.26785.01124}.



\bibitem{Horowitz:1999jd}
G.~T.~Horowitz, V.~E.~Hubeny, \href{https://doi.org/10.1103/PhysRevD.62.024027}{Phys. Rev. D \textbf{62}, 024027 (2000).}



\bibitem{Konoplya:2002zu}
R.~A.~Konoplya,
\href{https://doi.org/10.1103/PhysRevD.66.044009}{Phys. Rev. D \textbf{66}, 044009 (2002).}

\bibitem{Cardoso:2003cj}
V.~Cardoso, R.~Konoplya, J.~P.~S.~Lemos, \href{https://doi.org/10.1103/PhysRevD.68.044024}{Phys. Rev. D \textbf{68}, 044024 (2003).}

\bibitem{Wang:2004bv}
B. Wang, C. Y. Lin, C. Molina, \href{https://doi.org/10.1103/PhysRevD.70.064025}{Phys. Rev. D \textbf{70}, 064025 (2004).} 




\bibitem{Konoplya:2017ymp}
R.~A.~Konoplya, A.~Zhidenko, \href{https://doi.org/10.1103/PhysRevD.95.104005}{Phys. Rev. D \textbf{95}, 104005 (2017).}


\bibitem{Konoplya:2022hll}
R.~A.~Konoplya, A.~F.~Zinhailo, J.~Kunz, Z.~Stuchlik, A.~Zhidenko, \href{https://doi.org/10.1088/1475-7516/2022/10/091}{J. Cosmol. Astropart. Phys. \textbf{10}, 091 (2022).}

\bibitem{Fontana2022} R. D. Fontana, F. C. Mena, \href{https://doi.org/10.1007/JHEP10(2022)047}{J. High Energ. Phys. \textbf{2022}, 47 (2022).}  


\bibitem{Konoplya:2023kem}
R.~A.~Konoplya, A.~Zhidenko, \href{https://doi.org/10.1103/PhysRevD.109.043014}{Phys. Rev. D \textbf{109}, 043014 (2024).}




\bibitem{Ficek2024} F. Ficek, C. Warnick, \href{https://doi.org/10.1088/1361-6382/ad35a0}{Class. Quantum Grav. 41 085011 (2024).}  




\bibitem{Lin2024} J. Lin, M. Bravo-Gaete, X. Zhang, \href{https://doi.org/10.1103/PhysRevD.109.104039}{Phys. Rev. D \textbf{109}, 104039 (2024).}



\bibitem{Guo2024} Y. Guo, H. Xie, Y. G. Miao, \href{https://doi.org/10.1016/j.physletb.2024.138801}{Phys. Lett. B \textbf{855}, 138801 (2024).}









\bibitem{Caldwell2002} R. R. Caldwell, \href{https://doi.org/10.1016/S0370-2693(02)02589-3}{Phys. Lett. B \textbf{545}, 23 (2002).}

\bibitem{Nojiri2003} S. Nojiri, S. D. Odintsov, \href{https://doi.org/10.1016/S0370-2693(03)00594-X}{Phys. Lett. B \textbf{562}, 147 (2003).}

\bibitem{Elizalde2004} E. Elizalde, S. Nojiri, S. D. Odintsov, \href{https://doi.org/10.1103/PhysRevD.70.043539}{Phys. Rev. D \textbf{70}, 043539 (2004).}

\bibitem{Bronnikov2006} K. A. Bronnikov, J. C. Fabris, \href{https://doi.org/10.1103/PhysRevLett.96.251101}{Phys. Rev. Lett. \textbf{96}, 251101 (2006).}

\bibitem{Clement2009} G. Clément,  J. C. Fabris, M. E. Rodrigues, \href{https://doi.org/10.1103/PhysRevD.79.064021}{Phys. Rev. D \textbf{79}, 064021 (2009).}

\bibitem{Bronnikov2012} K. A. Bronnikov, R. A. Konoplya, A.  Zhidenko, \href{https://doi.org/10.1103/PhysRevD.86.024028}{Phys. Rev. D \textbf{86}, 024028 (2012).}


\bibitem{Jamil2011} M. Jamil, I. Hussain, M. U. Farooq, \href{https://doi.org/10.1007/s10509-011-0762-2}{Astrophy. Space Sci. \textbf{335}, 339 (2011).}

\bibitem{Pan2011} Q. Y. Pan, R. K. Su, \href{https://doi.org/10.1088/0253-6102/55/2/07}{Comm. Theor. Phys. \textbf{55}, 221 (2011).}




\bibitem{Jardim2012} D. F. Jardim, M. E. Rodrigues, S. J. M. Houndjo,  \href{https://doi.org/10.1140/epjp/i2012-12123-x}{Eur. Phys. J. Plus \textbf{127}, 123 (2012).}

\bibitem{Quevedo2016} H. Quevedo, M. N. Quevedo, A. Sanchez, \href{https://doi.org/10.1140/epjc/s10052-016-3949-4}{Eur. Phys. J. C \textbf{76}, 110 (2016).} 

\bibitem{Mo2018} J. X. Mo, S. Q. Lan, \href{https://doi.org/10.1140/epjc/s10052-018-6153-x}{Eur. Phys. J. C \textbf{78}, 666 (2018).}

\bibitem{Han2020} Y. W. Han, K. J. He, Y. Hong, \href{ https://doi.org/10.1007/s10773-020-04421-4}{Int. J. Theor. Phys. \textbf{59}, 1537 (2020).}


\bibitem{Salah} A. S. Mohamed, E. E. Zotos, \href{https://doi.org/10.1016/j.ascom.2024.100862}{Astron. Comput. \textbf{48},  100862 (2024).}

\bibitem{Shahzad2024} M. U. Shahzad, A. Mehmood, R. Gohar, A. \"Ovg\"un, \href{https://doi.org/10.1016/j.newast.2024.102225}{New Astron. \textbf{110}, 102225 (2024).}

\bibitem{Niki2009} P. Nicolini,  \href{https://doi.org/10.1142/S0217751X09043353}{Int. J. Mod. Phys A \textbf{24}, 1229 (2009).}

\bibitem{Nicolini} P. Nicolini, A. Smailagic, E. Spallucci, \href{https://doi.org/10.1016/j.physletb.2005.11.004}{Phys. Lett. B \textbf{632}, 547 (2006).}

\bibitem{Nozari} D. M. Gingrich, \href{https://doi.org/10.1007/JHEP05(2010)022}{J. High Energy Phys. \textbf{05}, 22 (2010).}

\bibitem{Tejeiro} J. M. Tejeiro, A. Larranaga, \href{https://doi.org/10.1007/s12043-011-0206-0}{Pramana J. Phys. \textbf{78}, 155 (2012).}

\bibitem{Rahaman} F. Rahaman, P. K. F. Kuhfittig, B. C. Bhui, M. Rahaman, S. Ray, U. F. Mondal, \href{https://doi.org/10.1103/PhysRevD.87.084014}{Phys. Rev. D \textbf{87}, 084014 (2013).}

\bibitem{Rizzo} T. G. Rizzo, \href{https://doi.org/10.1088/1126-6708/2006/09/021}{J. High Energy Phys. \textbf{09}, 021 (2006).}

\bibitem{Mehdipour} K. Nozari, S. H. Mehdipour, \href{https://doi.org/10.1088/0264-9381/25/17/175015}{Class. Quantum Grav.  \textbf{25}, 175015 (2008).}

\bibitem{Liang} J. Liang, B. Liu,  \href{https://doi.org/10.1209/0295-5075/100/30001}{Europhys. Lett. \textbf{100}, 30001 (2012).}

\bibitem{Hamil} B. Hamil, B. C. L\"{u}tf\"{u}o\u{g}lu, \href{https://doi.org/10.1016/j.dark.2024.101484}{Phys. Dark Universe \textbf{44}, 101484 (2024).}

\bibitem{Nicolini2005}  P. Nicolini, \href{https://doi.org/10.1088/0305-4470/38/39/L02}{J. Phys. A: Math. Gen. \textbf{38}, L631 (2005).}

\bibitem{Nasseri2005} F. Nasseri, \href{https://doi.org/10.1007/s10714-005-0183-z}{Gen. Relativ. Gravit. \textbf{37}, 2223 (2005).}

\bibitem{Campos2022} J. A. V. Campos, M. A. Anacleto, F. A. Brito,  E. Passos, \href{https://doi.org/10.1038/s41598-022-12343-w}{Sci. Rep. \textbf{12}, 8516 (2022).}

\bibitem{Mukherjee2008} P. Mukherjee, A. Saha, \href{https://doi.org/10.1103/PhysRevD.77.064014}{Phys. Rev. D \textbf{77}, 064014 (2008).}

\bibitem{Nozari2008} K. Nozari, B. Fazlpour, \href{https://www.actaphys.uj.edu.pl/fulltext?series=Reg&vol=39&page=1363}{Acta Phys. Polon. B \textbf{39}, 1363 (2008).}

\bibitem{Alavi2009} S. A. Alavi, \href{https://www.actaphys.uj.edu.pl/fulltext?series=Reg&vol=40&page=2679}{Acta Phys. Polon. B \textbf{40}, 2679 (2009).}

\bibitem{Nozari2010} K. Nozari, S. H. Mehdipour, \href{https://doi.org/10.1088/0253-6102/53/3/20}{Commun. Theor. Phys. \textbf{53}, 503 (2010).}


\bibitem{Ciric2025} M. D. Ćirić, N. Konjik, T.  Jurić,  A. Samsarov, I.  Smolić
Ivica Smolić, \href{https://doi.org/10.3390/sym17010054}{Symmetry \textbf{17}, 54 (2025).} 



\bibitem{Kim2008} W. Kim, E. J. Son, M. Yoon, \href{https://doi.org/10.1088/1126-6708/2008/04/042}{J. High Energy Phys. \textbf{04}, 042 (2008).}

\bibitem{Gangopadhyay2012} S. Gangopadhyay, D. Roychowdhury, \href{https://doi.org/10.1142/S0217751X12500418}{Int. J. Mod. Phys. A \textbf{27}, 1250041 (2012).} 

\bibitem{Ciric2018} M. D. Ćirić, N. Konjik, A. Samsarov, \href{https://doi.org/10.1088/1361-6382/aad201}{Class. Quantum Grav. \textbf{35}, 175005 (2018).} 

\bibitem{Ciric2020} M. D. Ćirić, N. Konjik, A. Samsarov, \href{https://doi.org/10.1103/PhysRevD.101.116009}{Phys. Rev. D \textbf{101}, 116009 (2020).} 














\bibitem{Liang2017} J. Liang,  Z. H. Guan,  Y. C. Liu, B. Liu, \href{https://doi.org/10.1007/s10714-017-2189-8}{Gen. Relativ. Gravit. \textbf{49}, 29 (2017).} 




\bibitem{Hang} H. Liu, \href{https://doi.org/10.1140/epjc/s10052-023-12066-w}{Eur. Phys. J. C \textbf{83}, 935 (2023).}

\bibitem{Sunny} S. Vagnozzi et al., \href{https://doi.org/10.1088/1361-6382/acd97b}{Class. Quantum Grav. \textbf{40}, 165007 (2023).}  

\bibitem{Johnson} C. V. Johnson, \href{https://doi.org/10.1088/0264-9381/31/20/205002}{Class. Quantum Grav. \textbf{31}, 205002 (2014).}

\bibitem{dyoni} Kh. Jafarzade, J. Sadeghi \href{https://doi.org/10.1007/s10773-017-3501-9}{Int. J. Theor. Phys. \textbf{56}, 3387 (2017). } 

\bibitem{Gauss} C. V. Johnson,\href{https://doi.org/10.1088/0264-9381/33/21/215009}{Class. Quant. Grav. \textbf{33}, 215009 (2016).} 

\bibitem{Infeld} C. V. Johnson, \href{https://doi.org/10.1088/0264-9381/33/13/135001}{Class. Quantum Grav. \textbf{33}, 135001(2016).} 

\bibitem{dilatonic} C. Bhamidipati, P. Kumar Yerra, \href{https://doi.org/10.1140/epjc/s10052-017-5076-2}{Eur. Phys. J. C \textbf{77}, 534 (2017).}

\bibitem{BTZ} J. X. Mo, F. Liang, G. Q. Li, \href{https://doi.org/10.1007/JHEP03(2017)010}{J. High Energ. Phys. \textbf{2017}, 10 (2017).}

\bibitem{polytropic} M. R. Setare, H. Adami, \href{https://doi.org/10.1007/s10714-015-1979-0}{Gen. Relativ. Gravit. \textbf{47}, 133 (2015).}


\bibitem{Javed2025} F. Javed, A. Waseem, P. Channuie, G. Mustafa, T. Muhammad, E. G\"udekli, \href{https://doi.org/10.1016/j.dark.2024.101766}{Phys. Dark Universe \textbf{47}, 101766 (2025).}








\end{thebibliography}
\end{document}